\documentclass[aps,prb,superscriptaddress,showpacs,showkeys]{revtex4}

\usepackage{amsmath}
\usepackage{amsfonts}
\usepackage{bm}
\usepackage{graphicx}
\usepackage{subfigure}
\usepackage{color}

\newcommand {\Sm}{\mbox{${\mathbf{\cal{S}}}$}}
\newcommand {\si}{\mbox{\boldmath$\sigma$}}

\begin{document}

\title{Fano Regime of Transport through Open Quantum Dots}

\author{E.\ R.\ Racec}
\email{roxana@physik.tu-cottbus.de}
\affiliation{Technische Universit\"at Cottbus, Fakult\"at 1, Postfach 101344,
             03013 Cottbus, Germany }
\affiliation{University of Bucharest, Faculty of Physics, PO Box MG-11,
             077125 Bucharest Magurele, Romania}

\author{U. Wulf}
\email{wulf@physik.tu-cottbus.de}
\affiliation{Technische Universit\"at Cottbus, Fakult\"at 1, Postfach 101344,
             03013 Cottbus, Germany }

\author{P. N. Racec}
\email{racec@wias-berlin.de}
\affiliation{Weierstra\ss -Institut f\"ur Angewandte Analysis und Stochastik,
Mohrenstr. 39, 10117 Berlin, Germany }
\affiliation{National Institute of Materials Physics, PO Box MG-7,
             077125 Bucharest Magurele, Romania}

\begin{abstract}
We analyze a quantum dot strongly coupled to the conducting leads
via quantum point contacts - Fano regime of transport - and report a variety
of resonant states which demonstrate 
the dominance of the interacting resonances
in the scattering process in a low confining potential.
There are resonant states similar to the eigenstates of the isolated dot,
whose widths increase with increasing the coupling strength to 
the environment, and hybrid resonant states. The last ones
are approximatively obtained as a linear combination of 
eigenstates with the same parity in the lateral direction, 
and the corresponding resonances show 
the phenomena of resonance trapping or level repulsion.
The existence of the hybrid modes 
suggests that the open quantum dot behaves in the Fano regime like
an artificial molecule.
\end{abstract}


\maketitle

\section{Introduction}

In the past decades  quantum dots were among the most studied systems
in the solid state physics\cite{ferry,nazarov09}. Proposed in the eighties
as a system to minimize losses in the optical fiber\cite{khitrova}, 
quantum dots have currently become a subject for fundamental research
as artificial atoms\cite{khitrova}.
Here the typical properties of an isolated natural atom
\cite{kastner,ashoori96,ferry} are qualitatively reproduced 
even in the presence of the interaction with the environment \cite{goeres}.

An artificial atom is a system created in a semiconductor heterostructure
consisting of few electrons isolated from the environment by tunable
barriers\cite{khitrova,ferry}. 
These non-infinite barriers allow for attaching conducting 
leads that {\it open} the quantum dot for transport 
whereas the properties of the isolated dot survive to a certain degree 
depending on the coupling strength.
From the mathematical point of view,
the quantum system admits on each transport 
direction a continuous energy spectrum with resonances
instead of a set of discrete eigenenergies. 
In the case of very high barriers (weak coupling) 
the discrete energy levels  of the isolated system turn into
quasi-bound states. These are isolated resonances with an
extremely small imaginary part, i. e., long life-time, whose real part can be
well approximated by the eigenenergies
of the isolated dot \cite{brouwer01}. 
At weak coupling the physics
of the transport phenomena is dominated by the electron-electron 
interaction that induces a shift of the resonant states\cite{vasanelli02}
and the quantum dot follows the Coulomb blockade regime\cite{ferry,glazman02}. 
Decreasing the confinement barriers of the dot the coupling with 
the conducting leads increases and both, the tunneling phenomena 
and the spin interaction \cite{goldhaber-gordon,kalish05} become 
more and more important relatively to the Coulomb interaction.
In this intermediate regime the total transmission through the quantum
dot shows broad and slightly asymmetric peaks, the so-called Kondo resonances,
\cite{goldhaber-gordon,goeres} which are sensitive 
to the shape and the height of the confining barriers
\cite{vargiamidis}. 
Upon further decreasing of the
confining barriers the dot reaches the strong coupling  regime.
Here the total transmission through the quantum dot shows
asymmetric peaks and dips on a slowly varying
background\cite{goeres,kobayashi02,kobayashi03}.
These peaks exhibit a Fano line shape\cite{fano} 
with a complex asymmetry parameter.
They are narrower compared to the ones in the intermediate regime\cite{goeres}.
In the last years a number of studies
were reported considering various specific aspects
of transport in the strong coupling regime within non-interacting models
\cite{brouwer01,roxana01,racec02,levinson02,rotter03,ando04,satanin05,
vargiamidis,mendoza08}.
However, a satisfactory theory providing a complete description of the
scattering mechanisms in the low confinement potential\cite{goeres,kalish05}
at strong coupling does not exist. A particular difficulty,
in this type of scattering potential is the
high level density in the quantum system.
As known from the particle physics, the methods used
for describing quantum systems with a low level density, 
as the light atoms or nuclei,
are not applicable for heavy nuclei with a high level
density\cite{mueller09,rotter09}. 
For the mesoscopic physics, this means that
the scattering problem for
a quantum system in the strong coupling regime
requires a different treatment compared to
the scattering problem for a quantum dot in the Coulomb blockade regime.

In the strong coupling regime, the electron scattering is profoundly 
affected by the quantum interferences\cite{liu98}.
The indistinguishability of the identical quantum particles 
leads to the interference\cite{liu98} between electrons 
and consequently to the Fano effect\cite{fano,fano_book,levinson02}.
To explain this effect, 
often the existence of two 
interfering pathways or channels
is invoked, one of which is resonant while the other is non-resonant.
In the experiments in Ref. [\onlinecite{kobayashi02}]
there are two spatially well defined interference paths consisting of
the two arms of the Aharonov-Bohm ring \cite{kobayashi02,ando04}.
The arm in which a quantum dot is embedded defines the resonant path.
In the experiments of
G\"ores et al \cite{goeres} there
are no such clearly spatially separated interference paths,
and the understanding of the Fano effect in this case 
is not straightforward.
Clerk et al \cite{brouwer01} proposed as a nonresonant path a
trajectory directly connecting the source and the drain contacts and 
as the resonant one a path passing through the dot via a resonant state
and therefore spending a longer time in the dot. 
In this way, 
under certain conditions,
the complex asymmetry 
parameter of a single resonance is associated
with the dephasing time in the quantum system.
In the frame of this model, the quite narrow and strong asymmetric
("S-type" Fano) lines are found under the assumption that the quantum dot 
is coupled to two single-mode leads, but the slowly varying background
is not explained.
In Ref. [\onlinecite{satanin05}] another model is discussed, in which
the interfering paths are associated with 
the energy channels (subbands) of the leads: one of them contains
a resonance or a group of overlapping resonances at the energy 
of the incoming electron while the second one contains only 
propagating states at this energy. As the result of the interference, 
asymmetric Fano lines with a complex parameter are obtained.
In the presence of a scattering potential which couples the
two channels it is shown that an interaction between resonances 
corresponding to
different channels occurs, and this interaction exhibits dips
in the total transmission
for a favorable parity of the resonant states.
As a second effect of the coupling between the scattering channels,
the positions of the resonances corresponding to different channels
are strongly modified in the complex energy plane. In the strong coupling 
regime the information about the scattering channels 
is actually not relevant for understanding the interaction between 
resonances.

The above results  confirm our earlier idea\cite{roxana01} 
to define the interfering paths
using the resonances, i. e., the complex eigenvalues 
of the non-Hermitian Hamilton operator of the open system\cite{rotter03},
instead of the quantum numbers of the lateral problem in the leads 
(scattering channel numbers). 
The resonances are also  the singularities of the 
scattering matrix in the complex energy plane and, 
based on the decomposition
of the $S$-matrix in a resonant and a background term\cite{roxana01},
we have associated the interfering paths in a more formal way with 
these two terms.
In the limit of a quasi one-dimensional model
we have proven that the Fano function with a complex asymmetry 
parameter arises as the most general resonance line shape under 
the assumption that the background can be considered constant over 
the width of the resonance pole. The asymmetry parameter of the Fano line
reflects the strength of the interaction between the considered resonance 
and the background which contains the contributions of all 
other resonances.
These results were later confirmed in Refs. [\onlinecite{mendoza08,rotter03}].
For decoupled scattering channels the Fano lines are 
only slightly asymmetric. The strong asymmetric ones, like those found
by G\"ores et al \cite{goeres}, imply the existence of 
many channels in the leads which can be coupled by dint of
a nonseparable two-dimensional (2D) scattering 
potential\cite{satanin05,mendoza08}.
As mentioned
in Ref. [\onlinecite{satanin05}] the interaction between channels changes the
shape of the resonance dramatically.

In this paper we develop a resonant scattering theory that
takes properly into account
the mentioned high level density in the  
quantum system as well as a strong coupling of
the scattering channels in a nonseparable scattering potential.
In view of Ref. [\onlinecite{brouwer01}] we assume
that there exist direct trajectories which connect 
the source and the drain contacts,
i. e., the potential energy in the region of the point
contacts lies under the Fermi energy. 
According to the scanning electron microscope image of the 
system\cite{goeres}, the quantum point contacts are very short and the leads 
are wide, allowing for a few subbands. 
The number of the 
conducting channels in the source and drain contacts is essential 
for the coupling mechanism of the quantum dot to the contacts.
In the strong coupling regime and for a quantum system with a high
level density, they limit the number of the eigenstates which 
couple to the continuum. The other eigenstates become
consequently quasi-bound states\cite{mueller09,rotter09}.

We believe that, for a deep understanding of the transmission through 
a quantum dot in the strong coupling regime, 
a resonant theory for two-dimensional systems is 
indispensable. 
In our opinion a resonant perturbation 
theory on the base of the Feshbach formalism\cite{satanin05} is not sufficient 
for an accurate description of the strong coupling between the scattering 
channels. The resonances 
characterize the 2D scattering 
potential, and a direct solution of the 2D Schr\"odinger equation can 
not be avoided at least in the strong coupling regime. 
For this purpose we use here the R-matrix 
method\cite{wigner47,lane58,smrcka90,wulf98,roxana01,racec09}
and extend our scattering theory for 1D systems without spherical 
symmetry\cite{roxana01} to the case of 2D systems. The R-matrix 
formalism is a very powerful method which allows for an efficient 
procedure to determine the resonances and for an exact  decomposition of the 
scattering matrix into resonant and nonresonant contributions
around each resonance\cite{roxana01}.
The second advantage of using 
the R-matrix formalism is that the scattering theory 
can be extended
to describe the wave functions inside the scattering area. 
In this way the electron 
probability distribution density within the dot region can be analyzed,
and the resonant states can be  compared with the atomic orbitals.
The coupling between the scattering channels
leads to the occurrence of the hybrid resonant modes.
Similar hybrid modes have also been evidenced in rectangular
electromagnetic resonators\cite{yang07} yielding a coupled mode with
low radiation losses and a high Q-factor.
As their atomic orbital counterparts, for example
in $H_2O$-molecules, hybrid resonant modes
arise in response to external perturbations of the
isolated quantum system.
While for atoms this perturbation consists of the
molecular fields, for quantum dots the perturbation
is the interaction with the conducting leads.

\section{The model}
\label{model}

\begin{figure}[tb]
\begin{center}
\noindent\includegraphics[width=160mm]{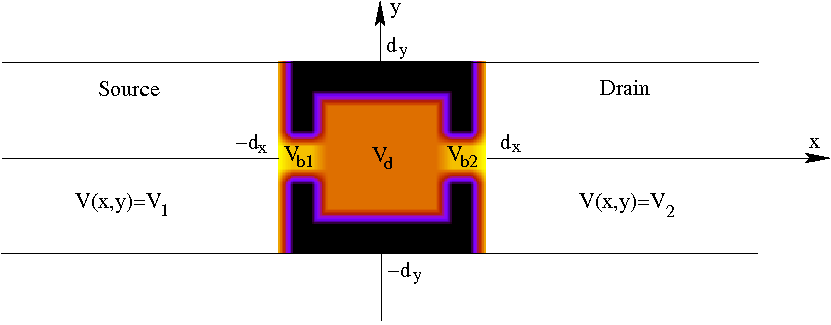}
\end{center}
\caption {(Color online)
Potential energy in the quantum wire: constant potential energy in the
source and drain contacts, $V_1$ and $V_2$, respectively, and position
dependent potential energy in the {\it scattering region} 
($|x| \le d_x$, $|y| \le d_y$). 
The quantum dot is isolated
inside the quantum wire by the barrier (black area) 
with the height $V_{b0}$ and the width $d_b$.
The coupling between dot and contacts is set by the potential energy
in the point contact regions, $V_{b1}$ and $V_{b2}$.
The constant potential energy felt by the electrons inside
the dot is $V_d$.
At the interface between different domains the potential 
energy varies linearly with position. 
}
\label{s_pot_1}
\end{figure}

We provide a model for transport through a quantum dot that can be
analyzed individually, like a single electron 
transistor\cite{kastner,goldhaber-gordon,goeres,kalish05}. 
The dot is embedded in a quite wide and infinitely
long quantum wire, isolated inside a two-dimensional electron gas (2DEG)
 by infinite
barriers, $V(x,|y| > d_y) \rightarrow \infty$. Inside the wire the dot
is defined by the barrier $V_b(x,y)$, which can have a general form
(nonseparable and without any symmetry) like the black area in  
Fig. \ref{s_pot_1}. At $y=0$ there are two quantum point contacts that ensure 
a strong coupling between the quantum dot and the rest of the wire, 
which plays the role of the source and drain contacts.
The contacts are characterized by constant potentials $V_s$
with $s=1$ for the source  and $s=2$ for the drain. 
In the middle of the dot region the potential energy
$V_d$  is constant and can be varied continuously
by a plunger gate. 
Further, we make the assumption that there are no bound-states in our
system, i. e., $\min[V_1,V_2]=\min[V(x,y)]$, $\forall x$, $|y|<d_y$.

The electronic wave functions are solutions of the
two-dimensional Schr\"odinger equation 
\begin{equation}
\left[ -\frac{\hbar^2}{2 m^*} 
        \left( \frac{\partial^2}{\partial x^2}
              +\frac{\partial^2}{\partial y^2}\
        \right)
      +V(x,y)
\right] \; \psi(x,y)
= E \; \psi(x,y),
\label{Sch1}
\end{equation}
with the general nonseparable potential $V(x,y)$ in the dot region; 
$E$ denotes here the kinetic energy of
the electron in the plane of 2DEG and $m^*$ its effective mass. 
The Fermi energy \cite{AFS82} for the 2D problem
is fixed by the electron density $N_S$ of 2DEG,
$E_F= \pi N_S \hbar^2/g_v m^*$, where
$g_v$ is the valley degeneracy factor.

The electronic transport through a single quantum dot 
is essentially a scattering
process\cite{buettiker86,buettiker85} for which the potential energy has a 
spatial dependence only within a quite small region of the structure 
called {\it scattering region} ($|x| \le d_x$, $|y| \le d_y$)
and is constant outside it. As usual in the scattering theory
\cite{newton} this type of problem is solved using different 
methods for these two  
regions of the structure 
and the solutions are connected 
based on the continuity
conditions of the wave function and its first derivative.

Outside the scattering region the Schr\"odinger equation 
is exactly solvable and, as usual in the scattering theory,
those solutions 
are
superpositions of \textit{one} incident and many scattered
waves, 
\begin{equation}
\psi_n^{(s)}(E;x,y)
= \frac{\theta(N_s(E)-n)}{\sqrt{2 \pi}}
  \begin{cases}
     \delta_{s1} \exp{[ i \; k_{1n} \; (x+d_x)]} \; \phi_n(y)
    & \cr
    +\sum_{n'=1}^{\infty}\limits \Sm^T_{sn,1n'}(E) 
                          \exp{[-i \; k_{1n'} \; (x+d_x)]} \; \phi_{n'}(y),
    & x \le -d_x \cr
     \delta_{s2} \exp{[-i \; k_{2n} \; (x-d_x)]} \; \phi_n(y)
    & \cr
    +\sum_{n'=1}^{\infty}\limits \Sm^T_{sn,2n'}(E) 
                          \exp{[ i \; k_{2n'} \; (x-d_x)]} \; \phi_{n'}(y),
    & x \ge d_x 
  \end{cases}
\label{psi_outside}
\end{equation}
$n \ge 1$, $s=1,2$, where $\theta$ denotes the  step function, 
i. e., $\theta(t)=1$ for $t \le 0$ and $\theta(t)=0$ for $t < 0$,
and $\delta_{ss'}$ the Kronecker delta symbol,
i. e., $\delta_{ss'}=1$ for $s = s'$ and $\delta_{ss'}=0$ for $s \ne s'$.
The solutions (\ref{psi_outside}) of the Schr\"odinger equation
are called \textit{scattering functions}
and the matrix $\Sm$ is the {\it generalized scattering matrix}\cite{schanz95}
or the {\it wave transmission coefficients matrix}.
This is an infinite dimensional matrix,
which connects incoming and outgoing components of the
wave functions. $T$ denotes here the transpose matrix.
Due to the electron confinement in the infinite quantum 
well in the $y$-direction 
the functions $\phi_n(y)$ are given as
\begin{equation}
\phi_n(y) = \frac{1}{\sqrt{d_y}}
             \sin{\left[\frac{\pi n}{2 d_y}(y+d_y)\right]},
\quad n \ge 1
\label{phi}
\end{equation} 
and the corresponding eigenenergies are
\begin{equation}
E_{\perp n} = \frac{\hbar^2}{2 m^*} 
              \left(\frac{\pi}{2 d_y}\right)^2 n^2.
\label{E_perp}
\end{equation}
The quantum numbers $n$ associated with the lateral problem 
define the energy channels for transport on each side of the
scattering area, 
the so-called \textit{scattering channels}. 
The wave vectors are defined for every channel $(sn)$ as
\begin{equation}
k_{sn}(E) = k_0 \sqrt{(E-E_{\perp n}-V_s)/u_0},
\label{ksn}
\end{equation}
where $k_0=\pi/2d_x$ and  
$u_0 = \hbar^2 k_0^2/2 m^*$.
In the case of the  conducting or open channels, 
$k_{sn}$ are positive real numbers, 
while for the non-conducting or closed channels 
they are
given from the first branch of the complex square root function, 
$k_{sn}=i |k_{sn}|$. 
Thus, the number of the conducting channels, $N_s(E)$, $s=1,2$, 
is a function of
energy, and, for a fixed energy $E$,
this is the largest value of $n$ which satisfies the 
inequality $E-E_{\perp n}-V_s \ge 0$ 
for a given value of $s$.
The scattering functions  exist only for the conducting channels.

In the limit of a very low potential, 
i. e., $V(x,y) \rightarrow 0$, the scattering functions
become plane wave corresponding to the free electrons.
In the presence of a scattering potential with a non-separable character
a plane wave incident onto the scattering area is reflected on every
channel - open or closed for transport - of the same side of the system
and transmitted on every channel - open or closed for transport -
on the other side, the probability of each process being related 
to the elements of the generalized scattering matrix $\Sm$.
The $\theta$ function in Eq. (\ref{psi_outside}) 
restricts the number of the elements with a physical meaning 
in $\Sm$ to $N_1(E)+N_2(E)$ columns for each energy $E$.

\begin{figure}[t]
\subfigure[]{\includegraphics*[height=35mm]{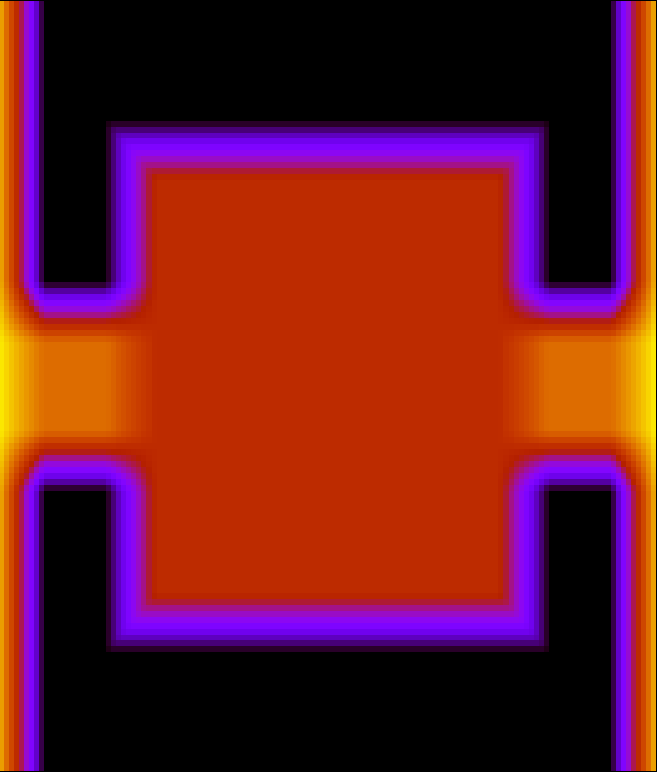}}
\subfigure[]{\includegraphics*[height=35mm]{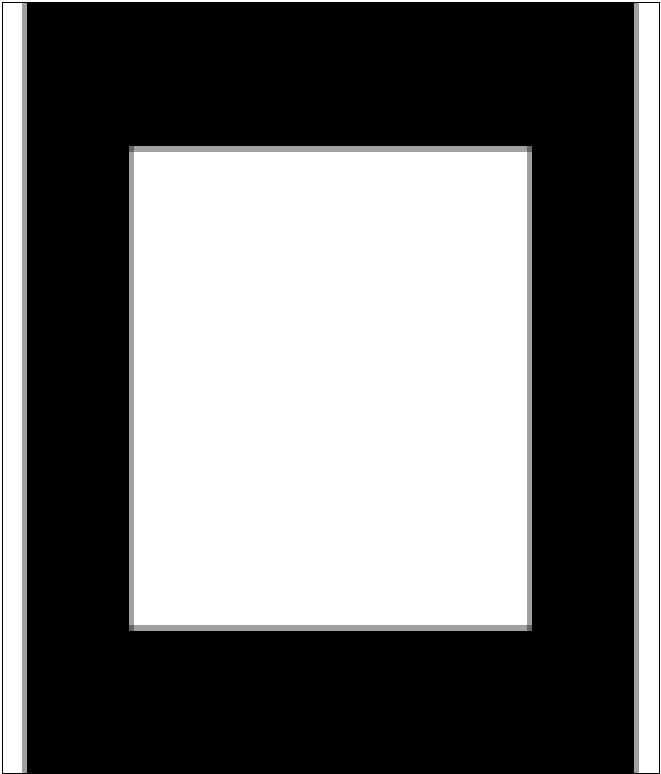}}
\caption{(Color online)
(a) Closed quantum dot by means of Neumann
boundary conditions at the dot-contact interfaces. 
This is the potential energy for the Wigner-Eisenbud problem.
(b) Isolated counterpart of the considered open quantum dot 
($V_b =V_{b1}=V_{b2} \rightarrow \infty$).
}
\label{dot}
\end{figure}

For further determining the generalized scattering matrix $\Sm$,
the Schr\"odinger equation  (\ref{Sch1}) should be also solved 
inside the scattering area.
In this domain the  
potential landscape does not generally allow for analytical solutions 
and we have chosen to solve Eq. (\ref{Sch1}) by means of
the R-matrix formalism\cite{wigner47,lane58}. 
Besides the extreme numerical efficiency\cite{racec09},
this powerful method allows 
for a direct comparison between  
the open quantum dot and its closed counterpart.
In the frame of the R-matrix formalism 
\cite{wigner47,lane58,smrcka90,wulf98,roxana01,racec09}, 
the scattering functions within the dot region, 
\begin{equation}
\psi_n^{(s)}(E;x,y) = \sum_{l=1}^{\infty} a_{ln}^{(s)}(E) \chi_l(x,y),
\label{psi_in_1}
\end{equation}
$|x| \le d_x$ and $|y| \le d_y$,
are expressed in terms of the eigenfunctions $\chi_l$
corresponding to the quantum dot artificially closed by Neumann boundary
conditions at the interfaces with the contacts,
\begin{equation}
\left. \frac{\partial \chi_l}{\partial x} \right|_{x=\pm d_x}=0,
\label{WE_cond}
\end{equation}
$l \ge 1$.
Thus, the so-called Wigner-Eisenbud functions $\chi_l(x,y)$ satisfy the 
same equation as $\psi_n^{(s)}(x,y)$,
Eq. (\ref{Sch1}), but  
with different  boundary conditions
in the transport direction: Since the scattering states 
$\psi_n^{(s)}(x,y)$ satisfy {\it scattering}, i.e. energy
dependent, {\it boundary conditions} 
derived from Eq. (\ref{psi_outside})
due to the continuity of the scattering functions at 
$x = \pm d_x$, the Wigner-Eisenbud function 
$\chi_l(x,y)$ has to satisfy energy independent boundary 
conditions given by Eq. (\ref{WE_cond}).
The infinite potential outside the quantum wire requires 
Dirichlet boundary condition 
on the surfaces perpendicular 
to the transport direction
for both functions,
$\psi_n^{(s)}(x,y=\pm d_y)=0$ and $\chi_l(x,y=\pm d_y)=0$.
The potential energy for the Wigner-Eisenbud problem is given 
in Fig. \ref{dot}(a).
As eigenfunctions of a Hermitian Hamilton operator
the functions $\chi_l$, $l \ge 1$, build a basis.
The corresponding eigenenergies are denoted by $E_l$
and are called Wigner-Eisenbud energies. They are real.

The expansion coefficients $a_{ln}^{(s)}(E)$ are calculated
using the Wigner Eisenbud eigenvalue problem and the boundary conditions
satisfied by the scattering functions at the interface with
the contacts. We have presented this method in detail 
in Ref. [\onlinecite{racec09}].
The coefficients  $a_{ln}^{(s)}(E)$ are obtained as a function of $\Sm$
and the scattering functions within the dot region
have the expression 
\begin{equation}
\vec{\Psi}(E;x,y) = \frac{i}{\sqrt{2 \pi}}
                    \mathbf{\Theta}(E) [{\bf 1} - \Sm^T(E)]
                    \mathbf{K}(E) \vec{R}(E;x,y)
\label{psi_in_2}
\end{equation}
with $\psi_n^{(s)}(E;x,y)=\left( \vec{\Psi}(E;x,y)\right)_{sn}$,
$n \ge 1$, $s=1,2$, and the R vector defined as
\begin{equation}
\vec{R}(E;x,y)= \frac{u_0}{\sqrt{k_0}}
                     \sum_{l=1}^{\infty}
                       \frac{\chi_l(x,y) \vec{\chi}_l}
                            {E-E_l}.
\label{R_vector}
\end{equation}
The vector $\vec{\chi}_l$ in Eq. (\ref{R_vector})
is constructed using the Wigner-Eisenbud functions
at $x=\pm d_x$ and the eigenmodes corresponding 
to the lateral problem in the contacts,
\begin{equation}
(\vec{\chi_l})_{sn} = \frac{1}{\sqrt{k_0}}
                       \int_{-d_y}^{d_y} dy \; \chi_l((-1)^{s} d_x,y)
                                               \phi_n(y),
\label{chi_vector}
\end{equation}
$n \ge 1, s=1,2$.
The wave vectors $k_{sn}$ define the diagonal matrix $\mathbf{K}$,
\begin{equation}
\mathbf{K}_{sn,s'n'}(E)= \frac{k_{sn}(E)}{k_0} \, \delta_{nn'} \delta_{ss'},
\label{K-matrix}
\end{equation}
$n,n' \ge 1$, $s,s'=1,2$.
The matrix $\mathbf{\Theta}$ is also a diagonal one, defined as
$\mathbf{\Theta}_{sn,s'n'}(E) = \theta(N_s(E)-n) \, 
                                 \delta_{ss'} \,
                                 \delta_{nn'}$, $n \ge 1$, $s=1,2$.

Using further the continuity of the scattering functions on the surface of
the scattering area 
one can derive
a relation between the $\mathbf{R}$-matrix 
\begin{equation}
\mathbf{R}(E)= u_0 \sum_{l=1}^{\infty}
                     \frac{\vec{\chi}_l \, \vec{\chi}_l^T}
		          {E-E_l}
\label{R-matrix}
\end{equation}
and the generalized scattering matrix $\Sm$,
\begin{equation}
\Sm(E) = \left[ \mathbf{1} - 2 \left( \mathbf{1} + i \mathbf{R} \mathbf{K} 
                                   \right)^{-1}
                \right] \mathbf{\Theta}(E).
\label{R-Srelation}
\end{equation}
The equation (\ref{R-Srelation}) is the key relation for solving 
2D scattering problems using only the eigenfunctions 
and the eigenenergies of the closed quantum dot [see Fig. \ref{dot}(a)]. 
They contain the information about the scattering potential
in the dot region and 
carry it over to the $\mathbf{R}$-matrix.
The matrix $\mathbf{K}$ characterizes
the contacts and can be constructed using only the constant values of 
the potential in these regions. 
On the base of  Eq. (\ref{R-Srelation}) the 
generalized scattering matrix $\Sm$
is calculated and further the scattering functions 
in each point of the
system are determined using
Eqs. (\ref{psi_outside}) and (\ref{psi_in_2}).
The scattering theory together with the R-matrix formalism 
allows for a complete description of the open quantum dot and 
each physical parameter of the system can be further derived from the 
$\Sm$-matrix.

According to Eq. (\ref{R-matrix}), $\mathbf{R}(E)$ is an 
infinite-dimensional symmetrical real matrix and its expression allows 
for a very efficient numerical implementation for computing it.
The big advantage of the R-matrix formalism is that, for a given potential 
landscape, only one eigenvalue problem with energy independent boundary 
conditions, i. e. Wigner-Eisenbud problem, has to be numerically solved and 
after that the generalized scattering
matrix $\Sm$ can be constructed for each energy using 
Eq. (\ref{R-Srelation}). The computational costs 
are in this case minimal, but most important is that Eq. (\ref{R-Srelation})
gives the explicit dependence of $\Sm$ on energy. This allows 
for an analysis of the scattering matrix and, after that, of the physical
properties of the system, in terms of resonance energies.

The generalized scattering matrix $\Sm$ describes the scattering processes
not only in the asymptotic region, but also inside the scattering area.
But this matrix is neither symmetric nor unitary
as can be seen from Eq. (\ref{R-Srelation}). 
Further, we define the \textit{current scattering matrix}
as
\begin{equation}
{\tilde{\Sm}} = \mathbf{K}^{1/2} \mathbf{\Theta} \Sm \mathbf{K}^{-1/2}.
\label{Stilde}
\end{equation}
The diagonal $\mathbf{\Theta}$-matrix in the above expression
ensure nonzero values only
for the matrix elements of $\tilde{\Sm}$ that correspond to conducting 
channels for which the transmitted flux is nonzero.
For simplicity,
we have dropped in (\ref{Stilde}) the energy dependence of the matrixes
and we will do this often henceforth.
Using the $R$-matrix representation of $\Sm$, Eq.
(\ref{R-Srelation}), the current scattering matrix becomes 
\begin{equation}
{\tilde{\Sm}} = \mathbf{\Theta}
                \left[ \mathbf{1} - 2 (\mathbf{1} + i \mathbf{\Omega})^{-1}
                \right]
                \mathbf{\Theta},
\label{Stilde2}
\end{equation}
with the symmetrical infinite matrix $\mathbf{\Omega}$
\begin{equation}
\mathbf{\Omega}(E) = u_0 \sum_{l=1}^{\infty}
                      \frac{\vec{\alpha}_l \, \vec{\alpha}^T_l}
                           {E-E_l}
\label{omega}
\end{equation}
and the column vector 
\begin{equation}
\vec{\alpha}_l(E) =\mathbf{K}^{1/2} \, \vec{\chi}_l,
\label{alpha}
\end{equation}
$l \ge 1$. According to Eq. (\ref{Stilde2})
the current scattering matrix ${\tilde{\Sm}}$
is also symmetric, ${\tilde{\Sm}}={\tilde{\Sm}}^T$.
The restriction of $\tilde{\Sm}$-matrix to the conducting channels is the well
known current transmission matrix \cite{wulf98,roxana01,racec09},
$\tilde{S}$,
commonly used in the
Landauer-B\"uttiker formalism. For a given energy $E$ this is a 
$(N_1+N_2) \times (N_1+N_2)$  matrix
which has to satisfy the unitarity 
condition $ \tilde{S} \tilde{S}^\dagger
            =\tilde{S}^\dagger \tilde{S} 
            ={\mbox{1}}$ 
according to the flux conservation\cite{schanz95}.

The elements of the current transmission matrix give directly
the reflection and transmission probabilities through the
quantum dot. For an electron incident from the contact $s=1,2$ 
on the channel $n$ the probability
to be transmitted into the contact $s' \ne s$ on the channel $n'$
is $T_{nn'}(E) = \left| {\tilde{\Sm}}_{2n',1n}(E) \right|^2 
               = \left| {\tilde{\Sm}}_{1n',2n}(E) \right|^2$.
In the case of non-conducting (evanescent) channels
these probabilities are zero.
With these considerations,
the total transmission through the dot, defined as
\begin{equation}
T(E) = \sum_{n=1}^{N_1(E)} \, \sum_{n'=1}^{N_2(E)} T_{nn'}(E),
\label{TT_def}
\end{equation}
becomes 
\begin{equation}
T(E) = \mbox{Tr} [\si(E) \si^\dagger(E)],
\label{TT}
\end{equation}
where $\si$ is the part of $\tilde{\Sm}$ 
which contains
the transmission amplitudes, $\si_{nn'}(E)=\tilde{\Sm}_{2n',1n}(E)$,
and $\si^\dagger$ its adjoint.

\section{Resonances}
\label{resonances}

The experimental analysis of a quantum system by coupling it to 
an electrical circuit has as a consequence the modification of 
its state. The physical interpretation of the measured
quantities cannot be based solely on the properties of 
the isolated quantum system, but rather on the properties of the 
open system, i. e., the quantum system coupled to the contacts. 
Due to this coupling, the eigenstates become resonant states, 
some of them are long-lived 
resonances (called simply resonances)
corresponding to  quasi-bound states\cite{ferry,rotter09}
and the other ones are practically delocalized\cite{ferry,rotter09}.
They can be found as states with a short life-time
and it has to be elucidated if they influence significantly  
the physical properties or not.
In addition, the resonant states are eigenstates of the non-Hermitian 
Hamilton operator\cite{mueller09,rotter09} 
of the open quantum system,
and they are not orthogonal to each other anymore\cite{mueller09,rotter09}.
In principle they can interact
and their coupling 
may also influence the physical properties.

From the mathematical point of view the resonances are associated 
with singularities of the current scattering 
matrix $\tilde{\cal{S}}$.
The representation of the $\tilde{\Sm}$-matrix in terms of 
$\mathbf{\Omega}$, Eq. (\ref{Stilde2}),
allows for a very fast and efficient numerical procedure 
to determine its poles. When the quantum dot becomes open, the
real eigenenergies of the closed system, $E_l$, migrate in the lower part
of the complex energy plane, becoming resonant energies\cite{bohm},
$\bar{E}_{0l} = E_{0l} - i \Gamma_l/2$, $l \le 1$. 
Based on this correspondence, 
we fix an energy $E_{\lambda}$ of the closed quantum  dot and determine the
resonance energy $\bar{E}_{0 \lambda}$ as a solution of 
the equation $\det[\mathbf{1} + i \mathbf{\Omega}(E)]=0$. 
The matrix $\Omega$, Eq. (\ref{omega}),
is split into a $\lambda$-dependent part and a rest
$\mathbf{\Omega}_{\lambda}$ which should be a slowly varying energy function
at least around an isolated resonance,
\begin{equation}
\mathbf{\Omega}(E) =  u_0 \frac{\vec{\alpha}_\lambda \, \vec{\alpha}^T_\lambda}
                     {E-E_\lambda} 
                  +\mathbf{\Omega}_\lambda(E).
\label{omega_lambda}
\end{equation}
As shown in Appendix A, this decomposition of $\mathbf{\Omega}$
around the resonance $\lambda$
leads to an expression of the $\tilde{\Sm}$ matrix 
in which the resonant and the background parts are
separated
\begin{equation}
{\tilde{\Sm}}(E) = 2 i u_0 \frac{\mathbf{\Theta} \, \vec{\beta}_\lambda 
                                 \,
                                 \vec{\beta}^T_\lambda \, \mathbf{\Theta}}
                                {E - E_\lambda -\bar{\cal{E}}_\lambda}
	           +{\tilde{\Sm}}_\lambda(E),
\label{Stilde3}
\end{equation}
where
\begin{equation}
\vec{\beta}_\lambda(E) = (\mathbf{1} + i \mathbf{\Omega}_\lambda)^{-1} 
                         \vec{\alpha}_\lambda
\label{beta}
\end{equation}
is an infinite column vector that characterizes the resonance $\lambda$,
\begin{equation}
\bar{\cal{E}}_\lambda(E) = -i \vec{\alpha}_\lambda 
                              \cdot
                              \vec{\beta}^T_\lambda 
\label{elambda}
\end{equation}
is a complex function which assures the analyticity of the current 
scattering matrix 
for every real energy and
\begin{equation}
{\tilde{\Sm}}_\lambda(E) = \mathbf{\Theta}
                       \left[ \mathbf{1} 
                             -2 (\mathbf{1} + i \mathbf{\Omega}_\lambda)^{-1}
                       \right]
                       \mathbf{\Theta}
\label{Stilde_lambda}
\end{equation}
is the background matrix.
We have already proposed in Ref. [\onlinecite{roxana01}] a decomposition 
of the scattering matrix similar to Eq. (\ref{Stilde3}), 
but for an effective one-dimensional scattering system 
without channel mixing.
There, the $\Omega$-matrix
is a $2 \times 2$ one and the inversion of $1+i \Omega$ reduces to
a simple algebraic calculation. In the presence of channel coupling
the inversion of an infinite matrix was a real mathematical challenge
(see Appendix \ref{poles}).

Based on the expression (\ref{Stilde3}) of the scattering matrix,
a resonant theory of transport through open quantum systems can be developed.
Eq. (\ref{Stilde3}) 
allows for the calculation of the resonance energies and for the
analysis of each resonant contribution to the conductance.
A similar decomposition of the transmission coefficient 
in a resonant term and a background is also proposed in Ref. 
[\onlinecite{mendoza08}], but in that case the two contributions can 
be evaluated
provided that the eigenvalue problem for the effective Hamiltonian 
of the open system is already solved.  

\begin{figure}[h]
\begin{center}
\noindent\includegraphics*[width=100mm]{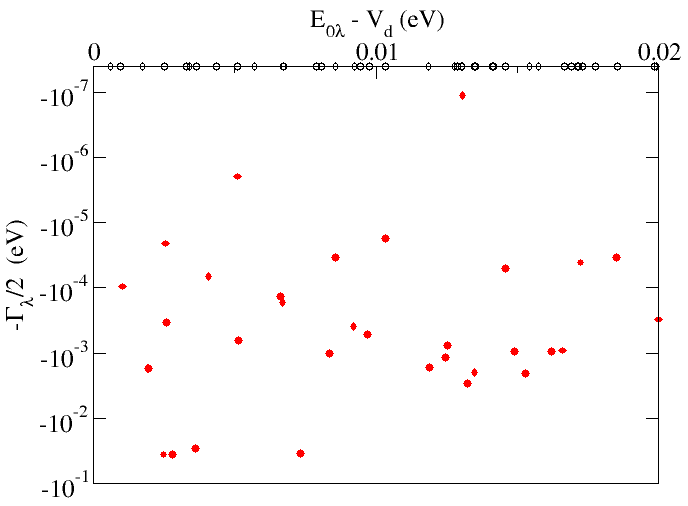}
\end{center}
\caption {(Color online) 
Resonance energies $\bar{E}_{0 \lambda}=E_{0 \lambda}-i \, \Gamma_\lambda/2$
(red filled symbols) of the
open quantum dot given in Fig. \protect\ref{s_pot_1}
with $V_d=0.0255124$ eV
and real eigenenergies $E_\lambda$ (black empty symbols)
of the corresponding closed dot. 
}
\label{fig-poles}
\end{figure}

Based on Eq. (\ref{Stilde3}), the position of the resonance 
$\bar{E}_{0 \lambda}=E_{0 \lambda} -i \Gamma/2$
in the complex energy plane is given as a
solution of the equation
\begin{equation}
 \bar{E} - E_\lambda - \bar{\cal{E}}_\lambda(\bar{E}) = 0,
\label{eq-poles}
\end{equation}
which can be solved numerically very fast using an iterative procedure
starting with $\bar{E}=E_\lambda$.
The complex function ${\bar{\cal{E}}}_\lambda(E)$, Eq. (\ref{elambda}), 
contains contributions from
all Wigner-Eisenbud energies and from all scattering channels, i. e., all
matrix elements of $\mathbf{K}$. Thus, the resonance energy 
$\bar{E}_{0 \lambda}$ can totally differ from $E_\lambda$
and only in the case of a very low coupling of the dot to the contacts, 
$E_\lambda$ can properly approximate 
the real part of the resonance energy $\bar{E}_{0 \lambda}$. 
With each resonance one can 
associate a resonance domain, which is a 
circle of radius $\Gamma_\lambda$ around
$\bar{E}_{0 \lambda}$ in the complex energy plane. 
The resonance energies for the quantum dot shown in Fig. \ref{s_pot_1}
are plotted in Fig. \ref{fig-poles} together
with the Wigner-Eisenbud energies.
The geometrical parameters used for the numerical calculations were 
taken from the electron micrograph of the single electron transistor
(SET) analyzed
in Ref. [\onlinecite{goeres}]:
$2d_x=2d_y=175$ nm, $d_b \simeq 35$ nm
so that the electrons are confined within a domain of 
about $100$ nm in diameter
and the point contact regions are about 35 nm $\times$ 35 nm.
The density of 2DEG is $N_S=8.1 \times 10^{11}$ cm$^{-2}$ 
and the Fermi energy $E_F=29,6$ meV.
The confining barrier was considered $V_{b0}=100$ meV
and in the point contact regions 
it was taken $V_{b1}=V_{b2}=2.5$ meV;
In the source and drain contacts $V_1 \simeq V_2 =0$.
At each interface between two domains the potential
energy varies linearly within a distance of 10 nm.
For the numerical calculation we fixed the number
of the scattering channels to $N_1=N_2=N_F=12$, where
$N_F$ is the number of the conducting
channels at the Fermi energy.
The analyzed scattering potential is not attractive
and, in turn, it is not expected that the evanescent channels play
an important role\cite{simon76}. Numerically, 
the inclusion of the evanescent channels
has not  produced significant variations of the conductance. 
In Ref. [\onlinecite{racec09}]
a detailed discussion is presented about the 
scattering potentials that allow for evanescent modes
and about the influence of these modes on the total transmission 
through an open quantum system.

As can be seen from Fig. \ref{fig-poles}, the  
open quantum dot strongly coupled to the source and drain 
contacts supports resonances with different widths, 
from very narrow, generally associated with modes localized
within the dot region, 
to very wide.
This phenomenon is known in the literature as 
{\it resonance trapping} \cite{mueller09,rotter09}:
only certain states of an open quantum system with overlapping 
resonances couple with the environment and their widths increase
with increasing the strength of the coupling, while the other ones
are more or less decoupled from the continuum\cite{mueller09,rotter09}. 
As illustrated in Fig. \ref{fig-poles}, for the quantum dot considered 
here there exist a few resonances with a long lifetime,
but the majority of the resonant
states couple to the contacts. 
This process is controlled by the number of the conducting
channels according to Refs. [\onlinecite{mueller09,rotter09}]:
In the strong coupling regime
$N_1(E)+N_2(E)$ resonant states couple to the environment
becoming quasi-delocalized.
As we have shown in Sec. \ref{model},
this number is energy dependent and increases with increasing $E$.
This result is physically correct
because the poles of the scattering matrix
having real part much higher than
the scattering potential, $E \gg \max[V(x,y)]$, have also a large 
imaginary part irrespective of the potential landscape.

The decomposition (\ref{Stilde3}) of the scattering matrix ${\tilde{\Sm}}$
in a resonant and a background term is especially 
relevant for energies
inside a resonance domain. 
According to Eq. (\ref{Stilde3})
all matrix elements of $\tilde{\cal{S}}$ 
and, in turn, all transmission coefficients $T_{nn'}$
between the scattering channels 
have a similar dependence on energy
around a resonance.  
\begin{figure}[h]
\begin{center}
\noindent\includegraphics*[width=100mm]{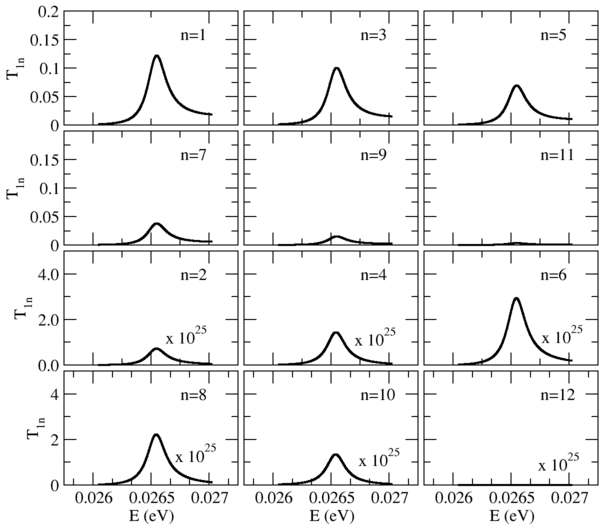}
\end{center}
\caption {Transmission as a function of energy around the 
resonance energy $E_{0 \lambda}=0.026537$ eV  
for the quantum dot described in Fig. \protect\ref{s_pot_1}
and $V_d=0.0255124$ eV.
}
\label{trans}
\end{figure}
In Fig. \ref{trans} we plot the transmission between the channel
$(11)$ and  the channels $(2n)$, $n \le N_F$
for energies around a fixed isolated resonance.
In the case of a symmetrical system the parity
plays an important role.
For the odd quantum number $n$ the function $\phi_n(y)$ has
the same symmetry as $\phi_1(y)$, and the transmission coefficient
$T_{1n}$ has a maximum around the resonance. If the parity is not
conserved the transmission is forbidden,
i. e., $T_{1n} \simeq 0$, $n=2,4,...$.
The plots in Fig. \ref{trans} confirm the similar 
energy dependence of the transmission coefficients and 
we can conclude that a resonance can be completely characterized
by the sum of these coefficients, i. e., by the
total transmission.
In Ref. [\onlinecite{ando04}] a similar idea was proposed
and a global Fano asymmetry parameter
was defined as a linear combination of the parameters 
corresponding to different scattering channels.

The plot of the transmission coefficients, Fig. \ref{trans},
shows a  strong coupling 
between the scattering channels in the Fano regime of transport.
The two quantum point contacts, specific for the SET geometry\cite{kastner,goeres},
control the strength of the coupling with the rest of the quantum wire
and confer the scattering potential its nonseparable character
responsible for the channel mixing.
In this case, a resonance cannot be associated anymore
with a single scattering channel as proposed in the models based 
on the Feshbach formalism in Refs. [\onlinecite{noeckel95,satanin05}]. 
The resonance perturbation theory\cite{satanin05} used there 
to describe the coupling between the
scattering channels can have limitations for 
large coupling strength and becomes certainly very 
laborious for a system with many conducting channels. In our model
the 2D Schr\"odinger equation is directly solved combining the
scattering theory with the R-matrix formalism and this method
can be used to describe each coupling regime.

\section{Conductance through open quantum dots}
\label{conductance}

The most common method to analyze experimentally a quantum dot is to
measure its conductance. 
In the limits of the Landauer-B\"uttiker 
formalism \cite{buettiker85,buettiker86} 
and for very low temperatures, the linear conductance is given as the 
total transmission through the dot at the Fermi energy,
\begin{equation}
G(V_d) = \frac{2 e^2}{h} T(E_F;V_d),
\label{G}
\end{equation}
for different values of the potential energy in
the dot region. Each variation of $V_d$ changes the scattering potential and,
in turn, the total transmission.

\begin{figure}[p]
\includegraphics[width=\textwidth]{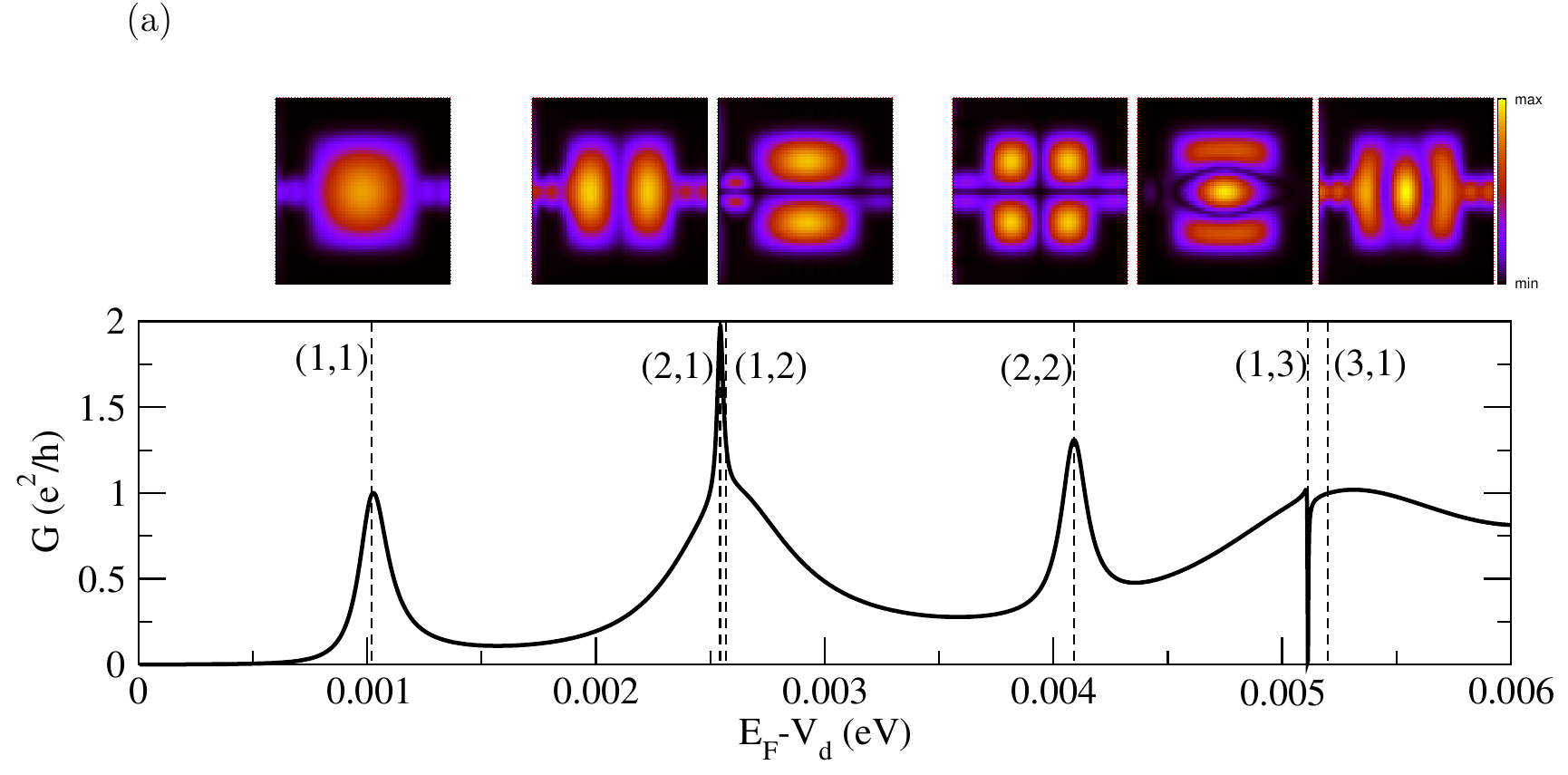}

\includegraphics[width=\textwidth]{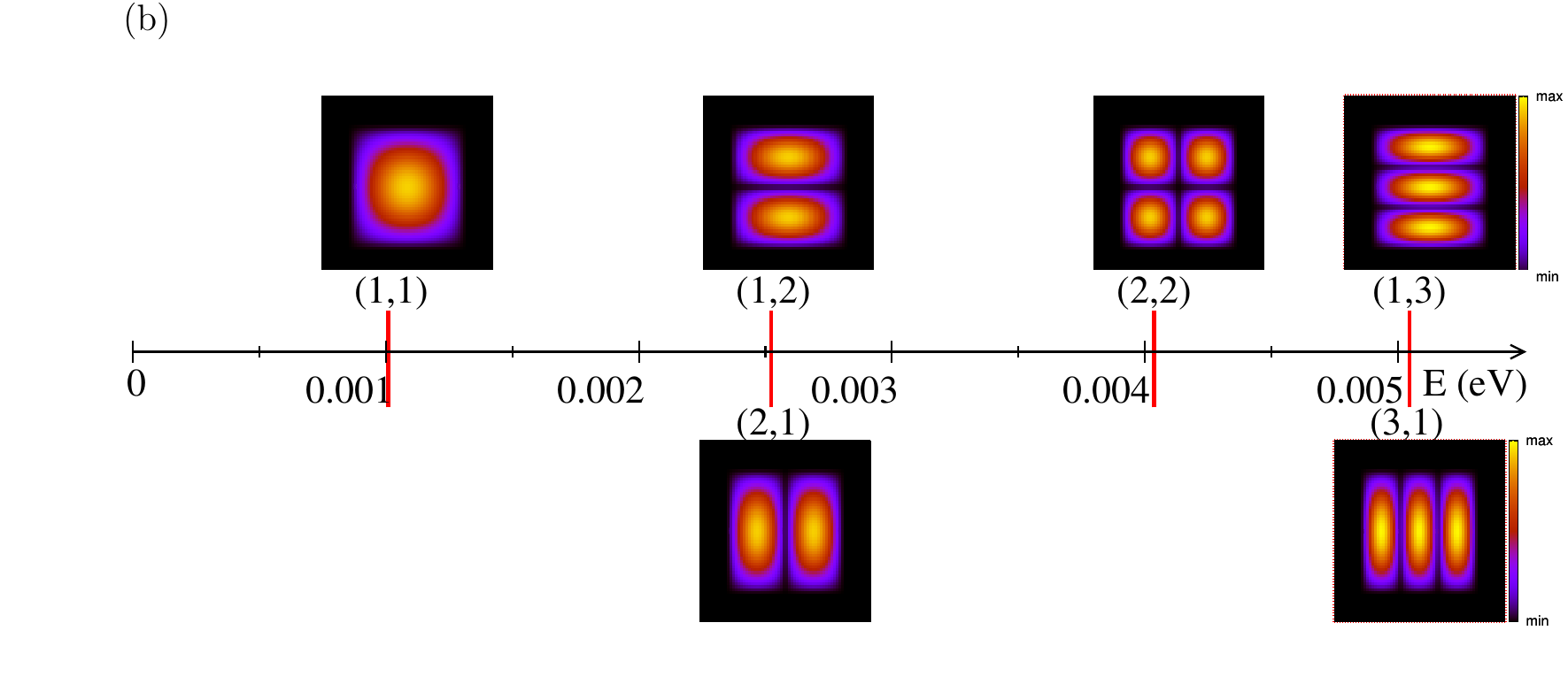}
\caption{(Color online)
(a):
Conductance as a function of the potential energy in the dot region.
The maps represent the 
electron probability distribution density inside the dot,
$|\psi_n^{(s)}(E_F;x,y)|^2$
for $V_d=V_0^{(n_x,n_y)}$ (vertical dashed lines)
for which $E_0^{(n_x,n_y)} \simeq E_F$;
The incident scattering channel is $n=1$ for the odd modes 
in the lateral direction and $n=2$ for the even ones.
(b):
The eigenenergies $\tilde{E}_{n_x,n_y}$ and the maps of the eigenstates,
$|\tilde{\psi}_{n_x,n_y}(x,y)|^2$,
for the isolated dot.
Bright 
corresponds to high values and dark corresponds to 
low values.}
\label{GG_fano_1}
\end{figure}

\begin{figure}[p]
\includegraphics[width=\textwidth]{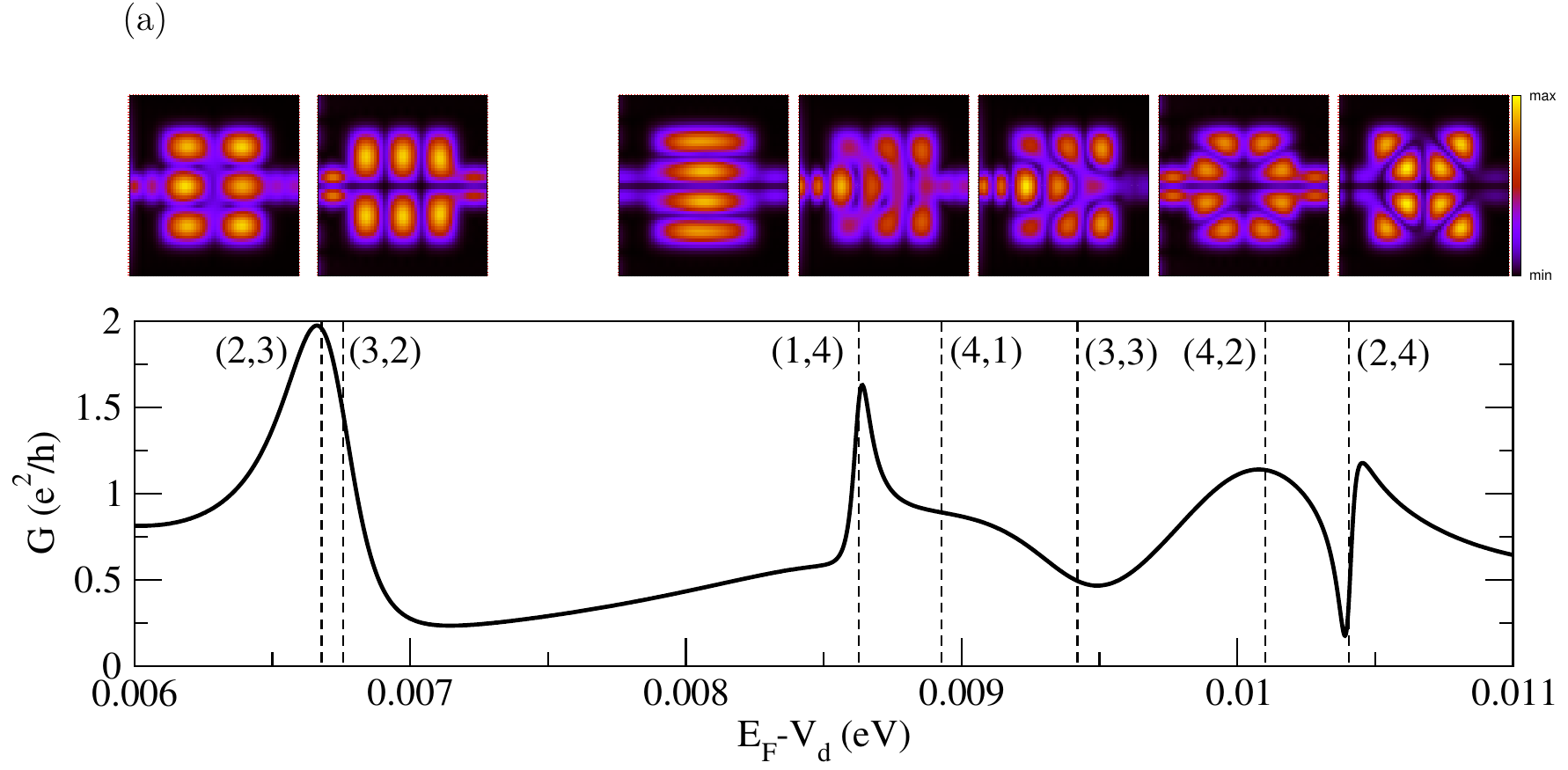}

\includegraphics[width=\textwidth]{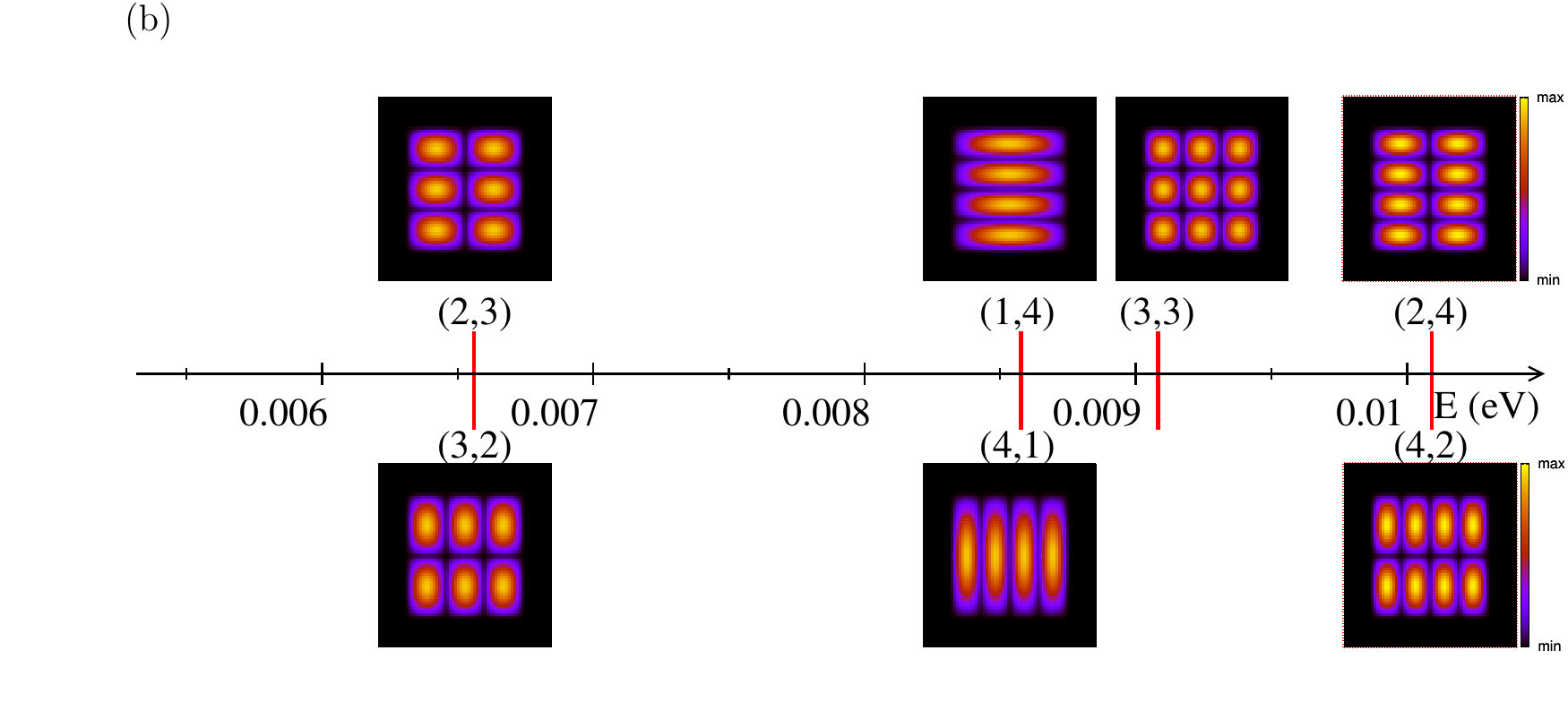}
\caption{(Color online)
(a):
Conductance as a function of the potential energy in the dot region.
The maps represent the
electron probability distribution density inside the dot,
$|\psi_n^{(s)}(E_F;x,y)|^2$
for $V_d=V_0^{(n_x,n_y)}$ (vertical dashed lines) 
for which $E_0^{(n_x,n_y)} \simeq E_F$;
The incident scattering channel is $n=1$ for the odd modes
in the lateral direction and $n=2$ for the even ones.
(b):
The eigenenergies $\tilde{E}_{n_x,n_y}$ and the maps of the eigenstates,
$|\tilde{\psi}_{n_x,n_y}(x,y)|^2$,
for the isolated dot.
Bright 
corresponds to high values and dark corresponds to
low values.}
\label{GG_fano_2}
\end{figure}

\begin{figure}[p]
\includegraphics[width=\textwidth]{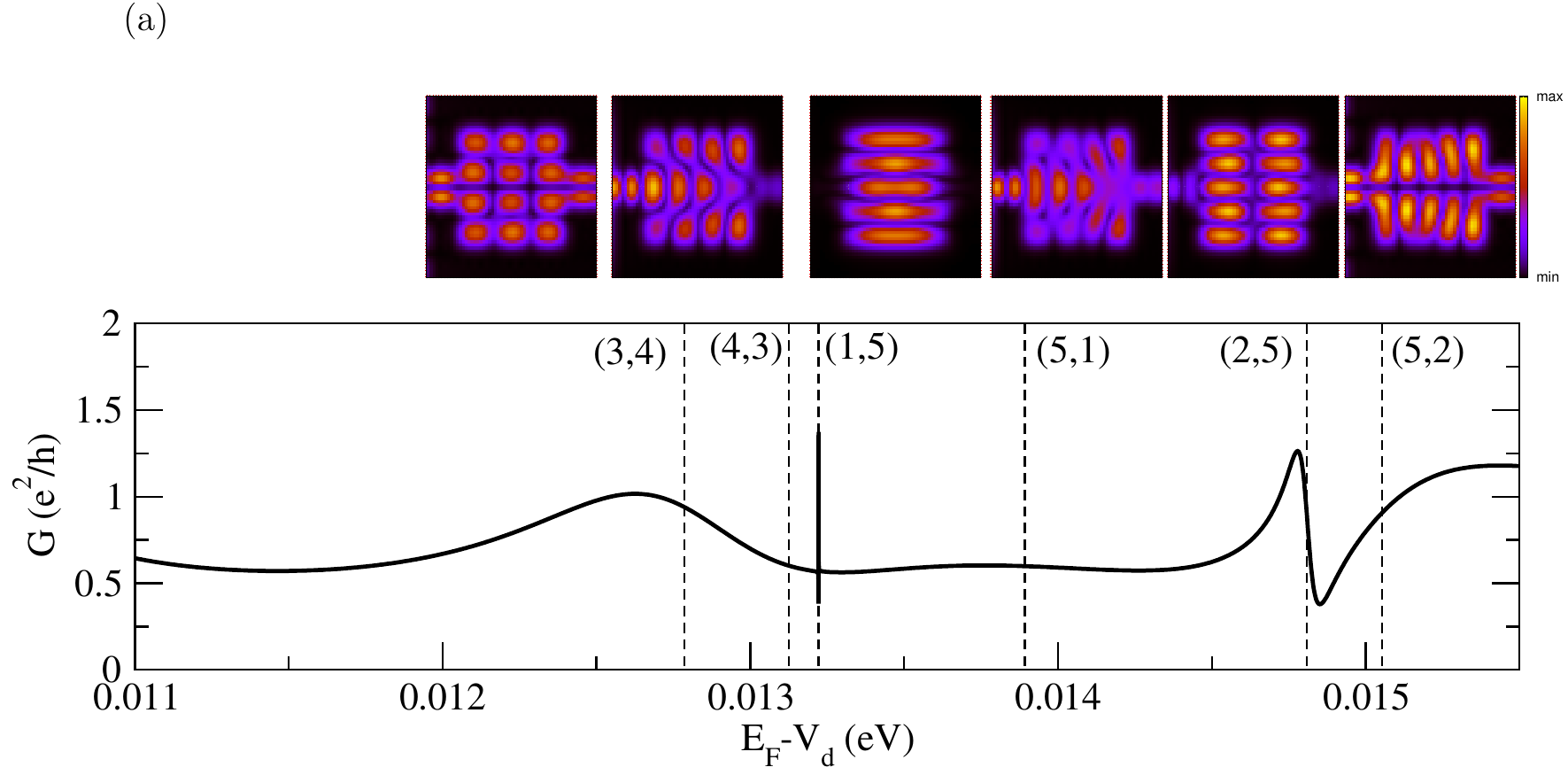}

\includegraphics[width=\textwidth]{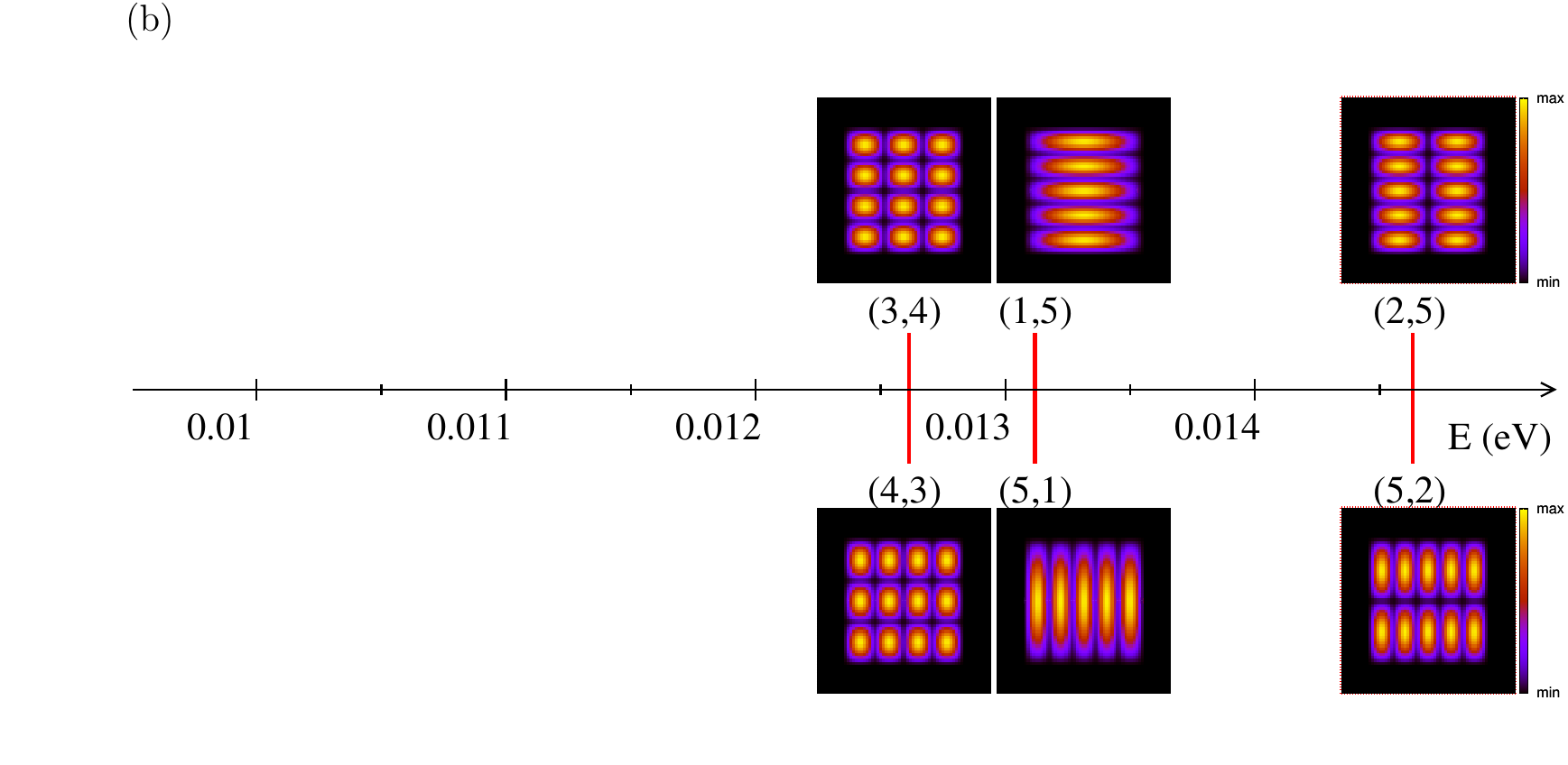}
\caption{(Color online)
(a):
Conductance as a function of the potential energy in the dot region.
The maps represent the
electron probability distribution density inside the dot,
$|\psi_n^{(s)}(E_F;x,y)|^2$
for $V_d=V_0^{(n_x,n_y)}$(vertical dashed lines)
for which $E_0^{(n_x,n_y)} \simeq E_F$;
The incident scattering channel is $n=1$ for the odd modes 
in the lateral direction and $n=2$ for the even ones.
(b):
The eigenenergies $\tilde{E}_{n_x,n_y}$ and the maps of the eigenstates,
$|\tilde{\psi}_{n_x,n_y}(x,y)|^2$,
for the isolated dot.
Bright 
corresponds to high values and dark corresponds to
low values.}
\label{GG_fano_3}
\end{figure}

In Figs. \ref{GG_fano_1}(a), \ref{GG_fano_2}(a), \ref{GG_fano_3}(a)
the conductance is plotted as a function of $E_F-V_d$
for the quantum dot presented in Fig. \ref{s_pot_1}
with the parameter given in Sec. \ref{resonances}.
The conductance shows peaks with line shapes from 
symmetric Breit-Wigner up to strong asymmetric ones
and even dips or antiresonances.
These maxima and minima are usually associated with resonances. 
Some peaks in the conductance reach values greater than 1 
and that means that at least two
resonances interplay to determine the line shape. 

We analyze further in detail, in terms of resonances, each type of 
peak and dip of the conductance.
For this purpose we need a functional dependence of the 
total transmission  on $V_d$, at least an approximation, 
around the peak maximum $V_0$. In the case of a dip in the conductance, 
$V_0$ denotes the position of the minimum. 
The R-matrix formalism used for solving 
the scattering problem allows, in a sense, for a very intuitive approach
of $T(E_F,V_d)$. 
A small variation $\delta V=V_d-V_0$ of the potential energy
felt by the electron in the dot region can be approximately 
seen as a shift of 
the potential energy in the whole scattering area. In turn, 
the Wigner-Eisenbud energies are shifted with $\delta V$ 
and the Wigner-Eisenbud functions remain unchanged. 
For the R-matrix, Eq. (\ref{R-matrix}), we can then write 
$ \mathbf{R}(E;V_0+\delta V) \simeq \mathbf{R}(E-\delta V;V_0)$.
This approximation is also valid for the total transmission
\begin{equation}
T(E_F;V_0 + \delta V)
\simeq
T(E_F-\delta V;V_0),
\label{T_app_F}
\end{equation}
because the wave vector $k_{sn}$ is a slowly varying energy function.
A detailed discussion about this approach is given in
Appendix A, Ref. [\onlinecite{roxana01}] for the open quantum dot
without channel mixing.
 
\clearpage

Based on the relation (\ref{T_app_F}),  each peak in
conductance can be associated with one or more resonances.
We consider first an isolated resonance with the complex energy 
$\bar{E}_{0 \lambda}=E_{0 \lambda}-i \Gamma_\lambda/2$.
The total transmission shows a peak around $E_{0 \lambda}$
that can be also seen in conductance if $E_{0 \lambda}$
matches the Fermi energy, $E_{0 \lambda}=E_F$.
Therefore, a maximum in conductance at $V_d=V_{0 \lambda}$ corresponds
to a resonance $\lambda$ and the quantity $E_F-V_{0 \lambda}$
gives the position of the resonance energy with respect to
$V_d$. In this way the resonance
energies can be directly compared with the eigenenergies of the isolated 
quantum dot. In view of the 
experiments presented in Ref. [\onlinecite{goeres}],
this is a square dot with the dimension $2d \times 2d$, $d=d_x-d_b$,
confined by a hard wall potential as depicted in Fig. \ref{dot}(b). 
Its eigenenergies 
\begin{equation}
\tilde{E}_{n_x,n_y} = V_d + \frac{\hbar^2}{2 m^*} 
                             \left( \frac{\pi}{2 d} \right)^2
                             (n_x^2 + n_y^2),
\quad n_x, \,\, n_y  \ge 1,
\label{eigenen}			     
\end{equation}
are plotted in Figs. \ref{GG_fano_1}(b), \ref{GG_fano_2}(b),
\ref{GG_fano_3}(b). 
The positions of the resonance energies are indicated by  
dashed lines in Figs. \ref{GG_fano_1}(a),
\ref{GG_fano_2}(a), \ref{GG_fano_3}(a).
As can be seen from these plots,
the open character of the quantum system
determines a shift of the eigenenergies in the complex energy plane, 
not only on the imaginary axis but also on the real axis. Due to the
two quantum point contacts, which couple the quantum dot to the source and drain,
the symmetry of the square dot is broken and the level degeneracy 
for $n_x \ne n_y$ is lifted. 

A deep understanding of the transport properties through
the open quantum dot requires a detailed analysis of the 
electron probability distribution density
within the dot region,
and the comparison of the resonance energies of the open dot
with the eigenenergies of the isolated dot\cite{ferry02}.
In the upper part of Figs. \ref{GG_fano_1}(a), \ref{GG_fano_2}(a),
and \ref{GG_fano_3}(a) the functions
$|\psi_n^{(1)}(E_F;x,y)|^2$, $n=1$ or $n=2$,
are given for $x$ and $y$ inside 
the scattering area and for $V_d$ corresponding to 
the maxima and minima in the conductance. 
These functions are called resonant states 
or resonant modes.
For comparison the eigenstates 
$|\tilde{\psi}_{n_x,n_y}(x,y)|^2$  of the 
isolated dot [see Fig. \ref{dot}(b)]
are presented in Figs. \ref{GG_fano_1}(b),
\ref{GG_fano_2}(b), and \ref{GG_fano_3}(b), where
\begin{equation}
\tilde{\psi}_{n_x,n_y}(x,y) = \frac{1}{d} 
                         \sin \left[ \frac{\pi n_x}{2d} (x+d) \right]
                         \sin \left[ \frac{\pi n_y}{2d} (y+d) \right],
\label{psi-tilde}
\end{equation}
$n_x \ge 1$ and $n_y \ge 1$.
The function $|\tilde{\psi}_{n_x,n_y}(x,y)|^2$ has
$n_x$ maxima in the $x$-direction and $n_y$ maxima in the $y$-direction.
All modes $(n_x,n_y)$ that we know from the isolated dot  
are also found for the open dot. Some of them are strongly 
modified due to the coupling with the contacts, but there are also modes 
that do not change much. 
Based on the similarities 
of the scattering functions at the resonance energy, 
$|\psi_n^{(s)}(E_F;x,y)|^2$ for $V_d=V_{0 \lambda}$ for
which $E_{0 \lambda}=E_F$, 
to the eigenfunctions
$|\tilde{\psi}_{n_x,n_y}(x,y)|^2$ of the isolated dot, we associate further
a pair of quantum numbers $(n_x,n_y)$ with each 
resonance $\lambda$, and
the resonance energies $\bar{E}_{0\lambda}$ will be further on denoted by
$\bar{E}_0^{(n_x,n_y)}=E_0^{(n_x,n_y)}-i \Gamma^{(n_x,n_y)}/2$. 
The potential energy in the dot region $V_{0\lambda}$, 
for which $E_0^{(n_x,n_y)}$
matches the Fermi energy, will be denoted by $V_0^{(n_x,n_y)}$.
In this way the resonances are classified using a very intuitive criterion.

\subsection{Peaks associated with isolated resonances}
\label{is_peaks}

First we analyze the slight asymmetric conductance peaks 
associated with an {\it isolated resonance} 
denoted by $\lambda$ 
or by $(n_{x},n_{y})$.
In the energy domain of this resonance the scattering matrix
$\tilde{\Sm}$ is given as a sum of a resonant term and a background,
Eq. (\ref{Stilde3}). Based on this relation and on the 
definition (\ref{TT}), 
the total transmission can be similarly decomposed.
According to the relation (\ref{T_app_F}) 
and for small variation $\delta V$ of the potential energy
around  $V_{0 \lambda}$ (for which $E_{0 \lambda} \simeq E_F$),
the conductance, Eq. (\ref{G}), follows the energy
dependence of the transmission 
and becomes 
\begin{equation}
G(V_{0 \lambda}+\delta V)
\simeq
G_{res}(E_F-\delta V;V_{0 \lambda})+G_{bg}(E_F-\delta V;V_{0 \lambda}).
\label{G_ir}
\end{equation}
The resonant contribution to the conductance is an energy dependent
function defined for each value of the potential energy in the dot 
region as
\begin{equation}
G_{res}(E;V_d) = \frac{2 e^2}{h} T_{0 \lambda}(E) 
         \left[ \left| \frac{2 i} {E - E_\lambda -\bar{\cal{E}}_\lambda(E)}
                      -\frac{1}{\bar{q}_\lambda(E)}
                \right|^2
               -\left| \frac{1}{\bar{q}_\lambda(E)} \right|^2
         \right]
\label{GC}
\end{equation}
with
\begin{equation}
T_{0 \lambda}(E) = \left| \vec{\beta}_{1 \lambda} \right|^2
                   \left| \vec{\beta}_{2 \lambda} \right|^2,
\label{T0_la}
\end{equation}
and the energy-dependent Fano asymmetry parameter\cite{rotter03} 
\begin{equation}
\frac{1}{\bar{q}_\lambda(E)}
= \frac{1}{T_{0 \lambda}}
  \vec{\beta}_{1 \lambda}^\dagger
  \si_\lambda
  \vec{\beta}_{2 \lambda}^{*},
\label{q_fano}
\end{equation}
where $(\vec{\beta}_{s \lambda})_{n} = (\vec{\beta}_\lambda)_{sn}$,
$s=1,2$ and $(\si_\lambda)_{n n'}= {\tilde{\Sm}}_{2n,1n'}$, $n,n' \ge 1$;
The symbol $*$ denotes the complex conjugate.
The background contribution to the conductance is given as
\begin{equation}
G_{bg}(E) = \frac{2 e^2}{h} 
            \mbox{Tr} [\si_\lambda(E)
                       \si_\lambda^\dagger(E)].
\label{GNC}
\end{equation}
The functions $G_{res}$ and $G_{bg}$ are obtained from 
the expression (\ref{Stilde3}) of the scattering matrix without any 
approximation.

The first contribution to the conductance, $G_{res}$, 
contains a resonant term singular at 
$E=\bar{E}_{0 \lambda}$ and a term 
$1/\bar{q}_\lambda$ that describes the coupling of the
resonance $\lambda$, characterized by the vector $\vec{\beta}_\lambda$, 
to the other resonances, characterized by the background matrix 
$\si_\lambda$. 
The function $G_{res}$  
yields always a peak mainly localized 
in the resonance domain. 
Due to the coupling 
of the considered resonance with the other ones,
this peak cannot have in principle a
Breit-Wigner line shape, even in the case of a narrow and isolated
resonance; 
The lowest approximation for a resonant peak is a  Fano
line shape with a complex asymmetry parameter 
obtained for $\bar{q}_\lambda(E) \simeq$ constant. 
The two terms add coherently to the
conductance and it is usual to call $G_{res}$ 
the coherent part \cite{roxana01}.
The second contribution to the conductance is the noncoherent 
part\cite{roxana01} given 
only by the background matrix $\si_\lambda(E)$.
In the case of an isolated resonance $\lambda$, it is expected 
that $G_{bg}$ is almost constant inside the resonance domain. 

The conductance curve given in Figs. \ref{GG_fano_1}(a), 
\ref{GG_fano_2}(a) and \ref{GG_fano_3}(a) shows two peaks that can be 
associated with isolated resonances. They correspond to the
resonances (1,1) and (2,2) as follows from the analysis of the  
electron probability distribution density in the dot region
[first and third maps in Fig. \ref{GG_fano_1}(a)].
For these two peaks 
the resonant and the background contributions to the conductance
are plotted in Fig. 8.
As expected, the resonant part is given by a slight asymmetric Fano 
line and the background is almost constant. 
\begin{figure}[h]
\subfigure[]{\includegraphics*[width=2.85in]{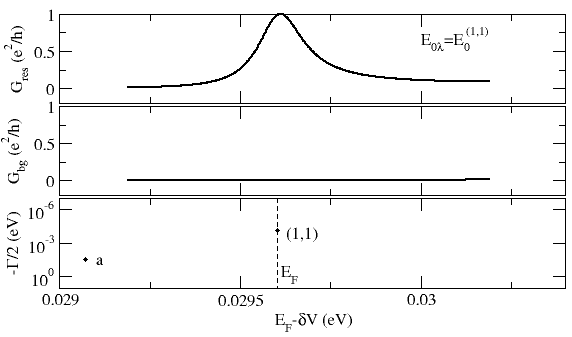}}
\hspace*{0.25cm}
\subfigure[]{\includegraphics*[width=2.85in]{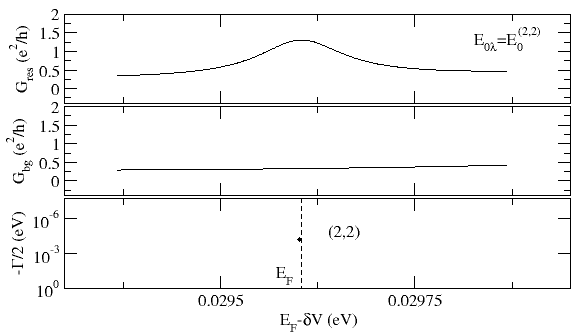}}
\caption{Conductance peaks associated with isolated resonances:
(a)$(n_{x},n_{y})=(1,1)$;
(b)$(n_{x},n_{y})=(2,2)$.
Upper part: Resonant part of the conductance $G_{res}$; 
Middle part: Background part of the conductance $G_{bg}$.
Lower part: Poles and position of the Fermi level 
in the complex energy plane.
The potential energy in the dot region is constant, 
$V_d=V_0^{(n_{x},n_{y})}$}.
\label{fano_1_1}
\end{figure}
But, unexpected is the fact that the two peaks are quite wide
compared to the other ones in the conductance curve.
The only possible explanation is related to the presence of the 
quantum point contacts, which modify dramatically the scattering process
and the picture which we have from the effective 1D scattering problem
is no longer valid. In the case of the quantum dot studied here,
the coupling between the scattering channels dominates the transmission 
through the dot and the scattering problem cannot be anymore reduced  
to a series of 1D problems. In turn, in the presence of 
the channel mixing the resonance widths  do not increase
monotonically with the energy.

\begin{figure}[h]
\subfigure[]{\includegraphics*[width=25mm]{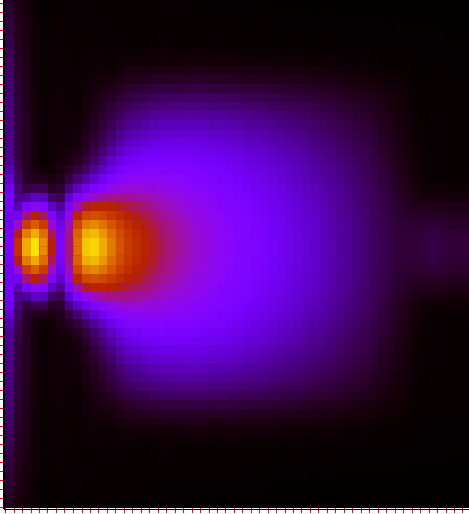}}
\subfigure[]{\includegraphics*[width=25mm]{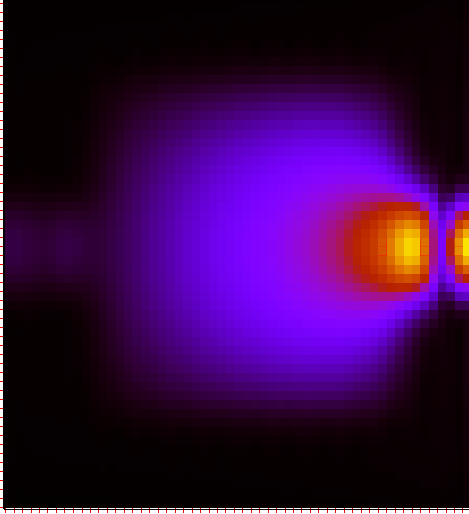}}
\subfigure[]{\includegraphics*[width=25mm]{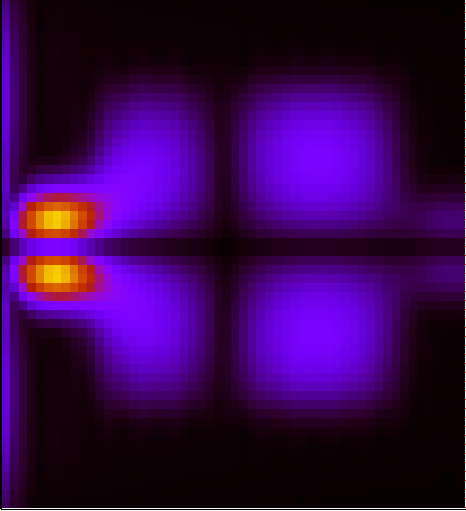}}
\subfigure[]{\includegraphics*[width=25mm]{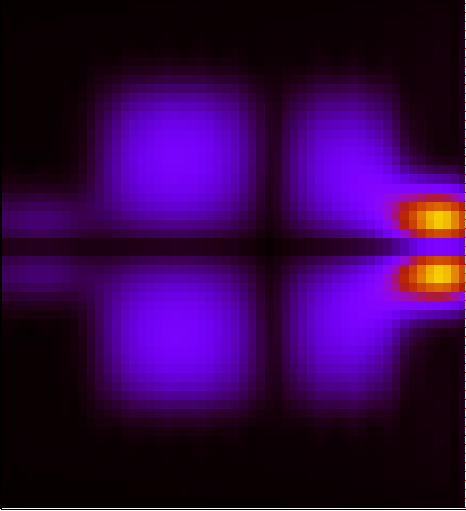}}
\caption{(Color online)
Electron probability distribution densities: 
(a) $|\psi^{(1)}_1(x,y)|^2$ and (b) $|\psi^{(2)}_1(x,y)|^2$
for the mode associated with the resonance "$a$" in Fig. \ref{fano_1_1}(a);
(c) $|\psi^{(1)}_2(x,y)|^2$, and (d) $|\psi^{(2)}_2(x,y)|^2$
for the mode associated with the resonance "$a$" in Fig. \ref{fano_1_1}(b).
}
\label{apertures}
\end{figure}

As can be seen in Fig. \ref{fano_1_1}, near the main resonances 
$(1,1)$ and $(2,2)$, there exist other ones denoted by "$a$".
They are broader, i. e., larger
imaginary part, and are associated with modes localized mainly
in the region of the two quantum point contacts, as shown in Fig. \ref{apertures}.
The resonances of type "$a$" do not influence directly 
the transmission through the open quantum dot, but they play 
a decisive role in the coupling process of an eigenstate 
to the continuum of states in the contacts. 
There are many modes localized in the point contact regions
with the resonant energies around the Fermi energy, but only for
a favorable symmetry they can intermediate a coupling between 
the quantum system and the source and drain contacts.
The probability distribution densities given in Fig. \ref{apertures}
shows evidently a coupling of the modes of type "$a$"
with the resonant modes $(1,1)$ and $(2,2)$. 
In this case, one can speak about an interaction between the two types
of resonances. In the next section we will study this phenomenon 
for resonances localized within the dot region, which are very close in
energy and have the same symmetry in the lateral direction.

\subsection{Peaks associated with overlapping resonances}
\label{ov_peaks}

Even in the case of a simple dot geometry, there exist only  a few 
isolated resonances. The other peaks in the conductance 
with strong asymmetric line shapes or maximum values about 2
are typical for the scattering processes dominated by 
two or more resonances, whose resonance domains cross, i. e.,
{\it overlapping resonances}.
How strong the overlapping resonances interact 
is determined by their relative position in energy\cite{ferry02} 
and by their
symmetry in the lateral direction\cite{satanin05} (perpendicular 
to the transport direction). 

Let assume that there exists a second resonance 
$\lambda' \equiv (n'_x,n'_y)$  
around the Fermi energy, i. e., in the vicinity of the first resonance
$\lambda \equiv (n_x,n_y)$, and this is a broader one,  
$\Gamma_\lambda' > \Gamma_\lambda$.
The presence of the second resonance leads 
to a strong variation of the background term $\tilde{\Sm}_\lambda$
with the energy around $E_F$. 
The expression of $\tilde{\Sm}_\lambda$, Eq. (\ref{Stilde_lambda}),
similar to $\tilde{\Sm}$, Eq. (\ref{Stilde3}),
allows for a further decomposition of this term
in a second resonant term and a new background.
Following the method described in Sec. \ref{resonances},
we can write
\begin{equation}
{\tilde{\Sm}}_\lambda(E) = 2 i u_0 \frac{\mathbf{\Theta} \, 
                                         \vec{\beta}'_\lambda \,
                                         \vec{\beta}'^{\,T}_\lambda \, 
                                         \mathbf{\Theta}}
                                         {E - E_{\lambda'} 
					  -\bar{\cal{E}}_\lambda'}
			 +{\tilde{\Sm}'}_\lambda(E),
\label{Stilde_lambda2}
\end{equation}
where $\vec{\beta}'_\lambda$, $\bar{\cal{E}}_\lambda'(E)$, and
${\tilde{\Sm}'}_\lambda$ can be obtained from 
$\vec{\beta}_\lambda$, $\bar{\cal{E}}_\lambda(E)$,
and ${\tilde{\Sm}}_\lambda$,
Eqs. (\ref{beta}), (\ref{elambda}), and (\ref{Stilde_lambda}),
respectively,
by replacing $\vec{\alpha}_\lambda$ by $\vec{\alpha}_{\lambda'}$
and $\mathbf{\Omega}_\lambda$ by  
$\mathbf{\Omega}'_\lambda = \mathbf{\Omega}_\lambda
                           -u_0 \frac{\vec{\alpha}_{\lambda'} \, 
		  	              \vec{\alpha}^T_{\lambda'}}
                                     {E-E_{\lambda'}}$.
Thus, the background contribution of the first resonance
to the conductance, Eq. (\ref{GNC}), 
becomes a sum of two contributions,
\begin{equation}
G_{bg}(E_F;V_d) = G'_{res}(E_F;V_d) + G'_{bg}(E_F;V_d),
\label{GNC2}
\end{equation}
a resonant one, 
\begin{equation}
G'_{res}(E;V_d) = \frac{2 e^2}{h} T'_{0 \lambda}(E) 
         \left[ \left| \frac{2 i} {E - E_{\lambda'} -\bar{\cal{E}}'_\lambda(E)}
                      -\frac{1}{\bar{q}'_\lambda(E)}
                \right|^2
               -\left| \frac{1}{\bar{q}'_\lambda(E)} \right|^2
         \right]
\label{GCp}
\end{equation}
with a similar energy dependence as $G_{res}(E;V_d)$
and a second background,
\begin{equation}
G'_{bg}(E;V_d) = \frac{2 e^2}{h}
                 \mbox{Tr} [\si_\lambda'(E)
	                    \si_\lambda'^\dagger(E)]
\label{GNCp}
\end{equation}
with $(\si'_\lambda)_{n n'}= {\tilde{\Sm}}'_{2n,1n'}$, $n,n' \ge 1$,
slowly varying with the energy if a third resonance does not
exist around $E_F$. 
The energy-dependent Fano asymmetry parameter $\bar{q}'_\lambda(E)$
in Eq. (\ref{GCp}),
associated with the resonance $\lambda'$, has the expression
\begin{equation}
\frac{1}{\bar{q}'_\lambda(E)}
= \frac{1}{T'_{0 \lambda}}
  \vec{\beta}_{1 \lambda}^{' \dagger}
  \si'_\lambda
  \vec{\beta}_{2 \lambda}^{'*},
\label{q_fano_p}
\end{equation}
where
$T'_{0 \lambda}(E) = \left| \vec{\beta}'_{1 \lambda} \right|^2
                     \left| \vec{\beta}'_{2 \lambda} \right|^2$
and $(\vec{\beta}'_{s \lambda})_{n} = (\vec{\beta}'_\lambda)_{sn}$,
$s=1,2$, $n \ge 1$.
The expression (\ref{GNC2}) is also
exact and we have only rearranged the terms  
in order to put directly in evidence
the contributions of each resonance
to the conductance. 
The second background term, $G'_{bg}$,
gives the possibility of a further decomposition in a third resonant 
term and a 
new background in the case of three
interacting resonances around the Fermi energy. 

For a systematic mathematical calculation we have also to consider
the energy dependence of $\si_\lambda$, Eq. (\ref{Stilde_lambda2}), in
the expression (\ref{q_fano}) of the Fano
asymmetry parameter associated with the first resonance,
\begin{equation}
\frac{1}{\bar{q}_\lambda}
= \frac{1}{T_{0 \lambda}}
  \left[ \vec{\beta}_{1 \lambda}^\dagger \si'_\lambda
         \vec{\beta}_{2 \lambda}^{*}
	+2 i u_0 \frac{ \vec{\beta}_{1 \lambda}^\dagger 
	                \cdot \vec{\beta}_{1 \lambda}' \,
	                \vec{\beta}_{2 \lambda}^\dagger
			\cdot \vec{\beta}_{2 \lambda}'}
  		      { E - E_{\lambda'} -\bar{\cal{E}}_\lambda'}
  \right].
\label{q_fano_2}
\end{equation}
The function $1/{\bar{q}_\lambda}$ is responsible for the asymmetry 
of the resonant contribution  $G_{res}$,
Eq. (\ref{GC}), that has a singularity at $\bar{E}=\bar{E}_{0 \lambda}$.
The presence of a second resonance $\lambda'$ around $\lambda$
yields in $1/{\bar{q}_\lambda}$ a term singular at 
$\bar{E}=\bar{E}_{0 \lambda'}$. 

If the first resonance is very narrow 
and the second one broaden, $\Gamma_\lambda \ll \Gamma_{\lambda'}$,
the Fano asymmetry parameter $1/{\bar{q}_\lambda}$ varies slowly
with the energy compared to the term in $G_{res}$ singular 
at $\bar{E}=\bar{E}_{0 \lambda}$. In this case, 
the energy dependence of $1/{\bar{q}_\lambda}$  can be neglected 
around the resonance $\lambda$ and an energy-independent 
Fano asymmetry parameter can be defined.
These results are in agreement with Ref. [\onlinecite{rotter03}].
The resonant contribution to the conductance is then given as a 
Fano function\cite{fano_book} $f(e)=|e+\bar{q}_F|^2/(e^2+1)$
with a complex asymmetry parameter $\bar{q}_F$.
For $|1/\bar{q}_F| \ll 1$ this function has a quasi Breit-Wigner profile,
while for $|1/\bar{q}_F| \gg 1$ it becomes a symmetric dip, usually called
antiresonance. The intermediate values $|1/\bar{q}_F| \simeq 1$
correspond to a Fano function characterized by a maximum and a minimum 
approximatively equidistant to the axis $e=1$ and we call this
profile a "S-type" Fano line. 
For open quantum dots, the different Fano profiles 
can be associated with different types of interacting resonances.
From Eq. (\ref{q_fano_2}) it follows  that 
$\bar{q}_\lambda \sim \vec{\beta}_\lambda \cdot \vec{\beta}'_\lambda$
for a second resonance much broader than the first one.
The two vectors $\vec{\beta}_\lambda$ and $\vec{\beta}'_\lambda$,
characterize the two considered resonances, and their scalar 
product is in principle nonzero. 
If the two resonant modes have different
parities in the lateral direction, 
the vectors $\vec{\beta}_\lambda$ and $\vec{\beta}'_\lambda$
are approximatively orthogonal to each other
and, in turn, the Fano asymmetry parameter 
has values from small to intermediate.
In this case, we can speak about a weak interaction between the 
overlapping resonances.
In contrast, for the same parity in the lateral direction,
the vectors $\vec{\beta}_\lambda$ and 
$\vec{\beta}'_\lambda$ are approximatively parallel to each other
and the Fano asymmetry parameter corresponds to a dip.
In this case, the two overlapping resonances interact strongly.
In Ref. [\onlinecite{satanin05}] the antiresonances in the conductance
through two identical quantum dots
embedded in a wave guide 
were also related to strong interacting resonances 
with the same parity.

In the case of two overlapping resonances with comparable widths,
both the term in $G_{res}$ singular at $\bar{E}=\bar{E}_{0 \lambda}$
and the Fano asymmetry parameter (\ref{q_fano_2}) vary 
slowly with the energy
and we can not predict the line shape around the resonances.
This situation corresponds to a wide peak in the conductance.

Summarizing all the above results and using the approximative
expression of the total transmission around a resonance at the 
Fermi energy, Eq. (\ref{T_app_F}), we obtain 
for the conductance 
\begin{equation}
G(V_{0 \lambda}+\delta V) 
\simeq 
  G_{res}(E_F-\delta V;V_{0 \lambda}) 
+ G_{res}'(E_F-\delta V;V_{0 \lambda}) 
+ G_{bg}'(E_F-\delta V;V_{0 \lambda}),
\label{G_2res_app}
\end{equation}
where $V_0^{(n_x,n_y)} = V_{0 \lambda}$ is the potential energy in the dot
region for which the resonance with the longest life-time
($\Gamma_\lambda < \Gamma_{\lambda'}$)
matches the Fermi energy.
Based on the above relation, we identify the contribution 
of each resonance to the conductance
and distinguish between weak and strong coupling regime of two
overlapping resonances.
The information about the
strength of the coupling between the resonances $\lambda \equiv (n_x,n_y)$
and $\lambda' \equiv (n'_x,n'_y)$ is contained into the energy-dependent
Fano asymmetry parameter, Eq. (\ref{q_fano}), and it determines the
line shape of the resonant contribution $G_{res}$ to the conductance.
The other two components of the conductance, $G'_{res}$
and $G'_{bg}$, provide information about the interaction
of the second resonance $(n'_x,n'_y)$ with all other resonances 
of the system excepting the two already considered.
A strong variation with the energy of the function $G'_{bg}$
in the energy domain of the resonance $(n'_x,n'_y)$ indicates the 
presence of a third resonance $(n''_x,n''_y)$ around the Fermi energy.
From the line shape of the resonant component $G'_{res}$
we can, in principle, get the information about
how strong this third resonance interacts with 
the resonance $(n'_x,n'_y)$.

\subsubsection{Weak interacting resonances}
\label{weak}

\begin{figure}[h]
\subfigure[]{\includegraphics*[width=2.85in]{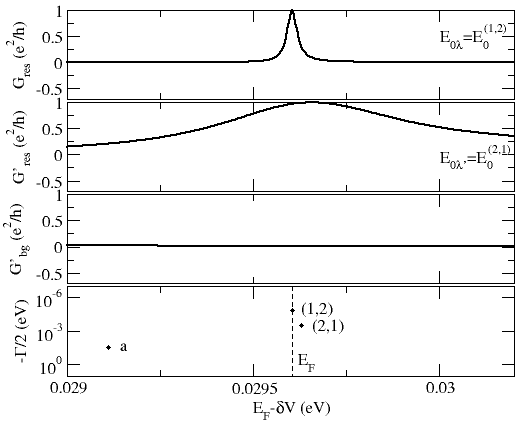}}
\hspace*{0.25cm}
\subfigure[]{\includegraphics*[width=2.85in]{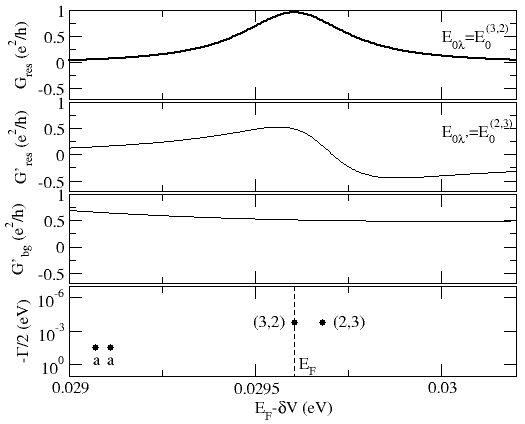}}
\caption{Conductance peaks associated with isolated resonances:
(a)$\lambda=(1,2)$, $\lambda'=(2,1)$;
(b)$\lambda=(2,3)$, $\lambda'=(3,2)$.
Upper parts: Resonant parts of the conductance $G_{res}$ and $G'_{res}$
and background part $G'_{bg}$;
Lower part: Poles and position of the Fermi level
in the complex energy plane.
The potential energy in the dot region is constant,
$V_d=V_{0 \lambda}$.}
\label{fano_1_2}
\end{figure}

In the weak interaction regime the two overlapping resonances
are close in energy but they do not perturb 
each other significantly. Each of them contributes to the conductance 
as a quasi-isolated resonance and the line shape of the peak 
is given as a superposition of two Fano lines
with a slight 
up to an intermediate asymmetry.
This is the case of the second peak in Fig. \ref{GG_fano_1}(a)
and the first peak in Fig. \ref{GG_fano_2}(a),
for which the different contributions to the conductance 
are  analyzed in detail  
in Fig. \ref{fano_1_2}.
The two peaks correspond to the pair of resonances $(1,2)$
and $(2,1)$ and $(2,3)$ and $(3,2)$.

In both situations the overlapping resonances 
have at the origin 
a degenerate eigenstate of the isolated dot presented in Fig. \ref{dot}(b),
with different symmetries in the $x$- and $y$-direction. 
The two quantum point contacts ($V_{b1}$ and $V_{b2}$)
of the open dot
create a strong coupling regime 
to the conducting leads and break the square symmetry of the isolated dot.
In turn, the degeneracy is lifted when the quantum dot becomes open
and, with increasing the 
coupling strength, the degenerate energy level 
evolves into two resonances that repulse
each other in the complex energy plane.
This phenomenon is illustrated in the lower 
part of Fig. \ref{fano_1_2}.
The first case corresponds to the 
{\it resonance trapping}\cite{mueller09,rotter09}
and the second one to the {\it level repulsion}\cite{mueller09,rotter09}.
Due to the trapping, the resonance $(1,2)$ has a longer life-time 
and a resonant state almost localized 
inside the dot region,
while the resonance $(2,1)$ has a shorter life-time 
and shows a significant probability
distribution density in the region of the two quantum point contacts.
The state with the nearest maximum to the aperture
couples stronger to the contacts and yields a broader contribution
to the conductance compared to the first resonance. 
Both of them are described by 
Fano lines with a slight asymmetry corresponding to a weak interaction
between the overlapping resonances and  with the background.
Figure \ref{fano_1_2}(b) shows the resonances 
$(2,3)$ and $(3,2)$, which have comparable widths and are well 
separated in energy. From the probability distribution density of these
two modes, Fig. \ref{GG_fano_2}(a), it is evident that both of them can 
easy couple to the modes localized in the point contact regions,
modes denoted by "$a$" and presented in Fig. \ref{apertures}.  
In this case, the resonance trapping is not favorable. 
The both resonant modes couple to the contacts via two modes of type "$a$". 
The line shape of $G_{res}$ 
corresponds to a weak interaction between the resonances $(2,3)$ and $(3,2)$,
but $G'_{res}$ indicates a stronger interaction of the second 
resonance with the background.
The total contribution to the conductance yields in each situation a 
peak with a maximum about 2. 

In conclusion, 
the weak coupling regime between overlapping resonances
is characterized by probability distribution densities within the dot region
similar to the eigenstates of the isolated dot.
According to this rule, the resonances (1,4) and (4,1) 
and (3,4) and (4,3) are also weak coupled with each other, but, as we
will see in the next section, in each case there is a strong coupling
with another neighbor resonance which modifies the probability 
distribution density of the states (4,1) and (4,3), respectively.
The last peak in Fig. \ref{GG_fano_3}(a) corresponds to the resonances 
$(2,5)$ and $(5,2)$ that interact also weakly 
and have a similar behavior to the pair $(2,3)$ and $(3,2)$.

\subsubsection{Strong interacting resonances. Hybrid modes} 
\label{strong}

The really new physics of the scattering process 
can be seen in the case of a strong interaction between 
the overlapping resonances,
phenomenon 
that does not occur 
in the case of an effective 1D quantum dot \cite{roxana01}.
This coupling regime 
is responsible for the thin or strong asymmetric
peaks and dips in the conductance 
and for the resonant states
whose probability distribution densities differ strongly from 
the corresponding eigenstates of the isolated dot.
Particularly for the SET geometry, Fig. \ref{s_pot_1},
the strong coupling of the quantum dot to the environment
is always accompanied by a strong scattering between the 
energy channels. This supplementary scattering determines 
the reordering process of the resonances 
in the complex energy plane, i. e., the
interaction between overlapping resonances. The
channel mixing influences especially the eigenstates 
with the same symmetry in the lateral direction.
Due to the favorable parity, these modes couple
with each other and generate new resonant modes that can not be
supported by the isolated dot.
As seen in Figs. \ref{GG_fano_1}, \ref{GG_fano_2}, and \ref{GG_fano_3},
there are two categories of strong coupled resonances:
The first ones are the resonances that correspond to eigenstates 
with the same symmetry in the $x-$ and $y-$direction
and whose resonant states are {\it hybrid modes}, 
similar to the hybrid orbitals
of the natural atoms. 
The resonances (1,3) and (3,1), (2,4) and (4,2), and (1,5) and (5,1) 
belong to this category.
The second category includes resonances 
corresponding to eigenstates with the same symmetry 
only in the lateral direction ($y$-direction)
like the pairs (4,1) and (3,3) and (4,3) and (5,1). 
We associate these resonances 
with a strong interaction because the probability distribution
densities for the states (4,1) and (5,1) are drastically modified 
in comparison with the isolated case. 

\begin{figure}[h]
\subfigure[]{\includegraphics*[width=2.85in]{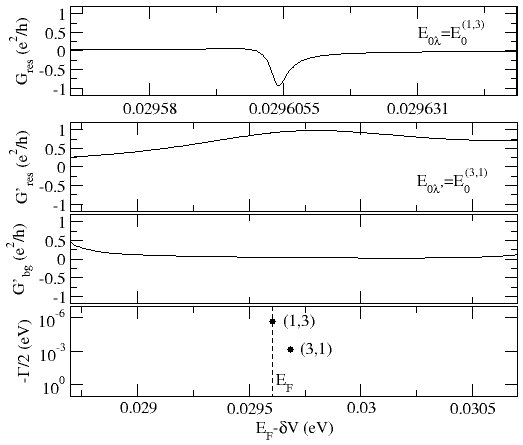}}
\hspace*{0.25cm}
\subfigure[]{\includegraphics*[width=2.85in]{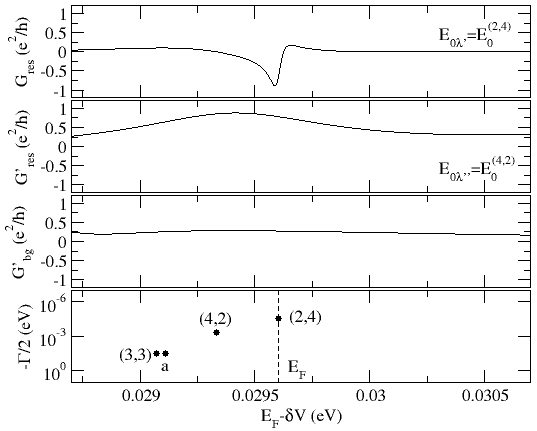}}
\caption{Conductance peaks associated with isolated resonances:
(a) $\lambda=(1,3)$, $\lambda'=(3,1)$;
(b) $\lambda=(2,4)$, $\lambda'=(4,2)$.
Upper parts: Resonant parts of the conductance $G_{res}$ and $G'_{res}$
and background part $G'_{bg}$;
Lower part: Poles and position of the Fermi level
in the complex energy plane.
The potential energy in the dot region is constant,
$V_d=V_{0 \lambda}$
}
\label{fano_1_3}
\end{figure}

The modes 
associated with strong interacting  resonances
yield dips or "S-type" Fano lines in the conductance,
superposed on the top of  broad peaks,
as shown in 
Figs. \ref{GG_fano_1}(a), \ref{GG_fano_2}(a), and \ref{GG_fano_3}(a).  
The overlapping resonance pairs
$(1,3)$ and $(3,1)$ and $(2,4)$ and $(4,2)$
are analyzed in detail in Fig. \ref{fano_1_3}.
The strong interaction of these resonances,
reflected by the dips in $G_{res}$, has as an effect
their strong repulsion in the complex energy plane.
In turn, the resonant modes $(1,3)$ and $(2,4)$ become long-lived
and are practically localized within the dot region, while
the modes $(3,1)$ and $(4,2)$ couple stronger to the contacts
and have shorter life-time. 
The electron distribution in the dot region favors in the
first case, Fig. \ref{fano_1_3}(a), a typical resonance trapping,
while in the second case, Fig. \ref{fano_1_3}(b), this phenomenon is 
not so pronounced, but it is 
accompanied by a level repulsion on the real axis.
The contribution of the second resonance to the conductance $G'_{res}$
is described by a broad peak and the background $G'_{bg}$ is almost constant.

Each pair of strong interacting resonances 
analyzed above corresponds to
a degenerate energy level of the isolated dot and their probability 
distribution densities are practically linear combinations of the two 
eigenfunctions of the degenerate level. This property can be easy seen 
in Fig. \ref{GG_fano_2}(a)
for the resonances $(2,4)$ and $(4,2)$. 
We can speak in this case about {\it hybrid resonant modes}.
The open quantum dot behaves like the oxygen atom 
in the water molecule: due to the interaction with the hydrogen atoms,
the $s$ and $p$ orbitals of the oxygen are mixed
to new hybrid orbitals so that the total energy of the molecule
is minimal. Similar, the coupling of the quantum dot to the contacts 
by means of two quantum point contacts 
yields a supplementary scattering potential
which allows for new resonant modes. They are not states which survive
the coupling process to the contacts \cite{ferry02}, but rather
new hybrid states, whose existence is directly connected to the
presence of the strong coupling regime. 
These modes 
offer the possibility of engineering quantum systems 
with complex properties.
Even in the case of a non-perfect square quantum dot the above results 
remain valid. A small difference between $d_x$ and $d_y$ yields instead 
of a degenerate level two very close eigenvalues. 
Essential for the strong interaction
of the two corresponding resonances is the same parity of the resonant states
on both directions and not the initial degeneracy.

\begin{figure}[h]
\subfigure[]{\includegraphics*[width=2.85in]{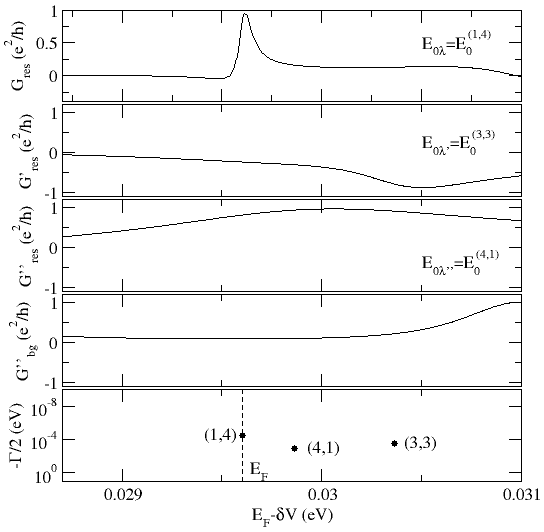}}
\hspace*{0.25cm}
\subfigure[]{\includegraphics*[width=2.85in]{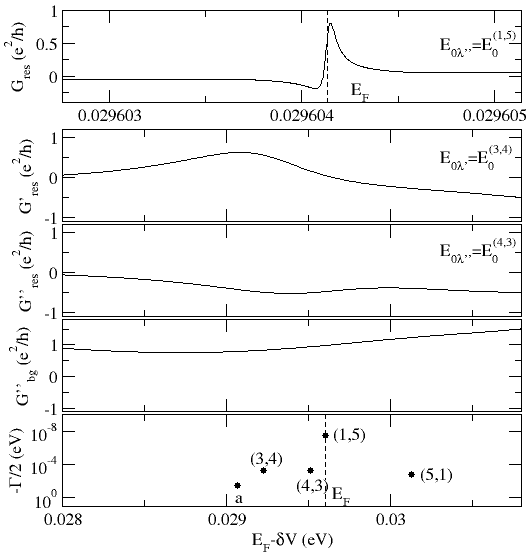}}
\caption{Conductance peaks associated with isolated resonances:
(a) $\lambda=(1,4)$, $\lambda'=(3,3)$, $\lambda''=(4,1)$;
(b) $\lambda=(1,5)$, $\lambda'=(3,4)$, $\lambda''=(4,3)$.
Upper parts: Resonant parts of the conductance $G_{res}$, $G'_{res}$,
and $G''_{res}$ and background part $G''_{bg}$;
Lower part: Poles and position of the Fermi level
in the complex energy plane.
The potential energy in the dot region is constant,
$V_d=V_{0 \lambda}$.
}
\label{fano_1_5}
\end{figure}

The resonances $(1,5)$ and $(5,1)$ interact also strongly.
They determine in the conductance 
a very thin "S-type" Fano line
superposed onto an extreme broad peak as shown in Fig. \ref{fano_1_5}(b). 
Their stronger repulsion in the complex energy plane 
compared to the  precedent cases (Fig. \ref{fano_1_3})
is determined by a supplemental strong interaction of the resonances
$(5,1)$ and $(4,3)$, which have the same parity in the
lateral direction.
The maps of the probability distribution densities
for the two modes in Fig. \ref{GG_fano_3}(a)
confirm also the phenomenon of hybridization.
The resonant modes $(4,3)$ and $(5,1)$ do not show such a high symmetry 
as the modes $(2,4)$ and $(4,2)$, but it is evident that they 
can be obtained as a linear combination of the eigenfunctions
$(4,3)$ and $(5,1)$ of the isolated dot and the mode $(4,3)$ 
dominates this combination.
The multiple interactions between neighbor resonances with the same
symmetry in the lateral direction amplify the phenomenon of
resonance trapping. One resonance - in this case resonance $(1,5)$ - 
decouples from the contacts and becomes extreme long-lived, 
while the other two 
become broaden and show a significant separation in energy.
If we consider the quantum dot as an artificial atom it is easy to accept
the hybridization as a natural process determined by the interaction with 
another system, but the existence of very narrow 
resonances supported by an open quantum dot
seems to be a paradox: it is necessary 
to open a quantum dot, i. e., to allow for regions where the direct electron 
transfer between dot and contacts is possible, in order to obtain
strongly localized states. Hence, long-lived modes of a quantum system 
can be obtained either in a quasi-isolated quantum dot or in a 
dot confined by shallow barriers engineered in such a way that
the scattering  channels are strongly mixed. 
The two types of localized modes 
have different fingerprints in the conductance:
in the first case they yield quasi-symmetric maxima, while in 
the second case strong asymmetric Fano lines appear on top of broad 
peaks.

The last sequence to be discussed 
corresponds to the resonances $(1,4)$, $(4,1)$, and $(3,3)$
in Fig. \ref{GG_fano_2}(a).
As can be seen from the probability distribution density maps
in Fig. \ref{GG_fano_2}(a),
there are three interacting resonances with different coupling strengths;
(1,4) and (4,1) interact weakly and yield two slight asymmetric maxima
in the conductance, one of them quite thin
and the other one broad, Fig. \ref{fano_1_5}(a).
In contrast, the resonances (4,1) and (3,3) interact strongly
and the resonant contribution of $(3,3)$ is
a broad dip.
The three interacting resonances $(4,1)$, $(3,3)$ and $(1,4)$
are very interesting in view of the experiments presented 
in Ref. [\onlinecite{goeres}]. Their contribution to the 
conductance together with the next peak
determined by the resonances $(2,4)$ and $(4,2)$,
Fig. \ref{GG_fano_2}(a),
approximate qualitatively very well the conductance curve
given in Ref. [\onlinecite{goeres}], Fig. 2(a),
for the quantum dot in the Fano regime.
Based on our resonance analysis we can conclude that 
the first thin peak in the measured conductance curve
is superimposed on the top of a second broad peak
and they correspond to two weak interacting 
resonances with different
symmetries in the lateral direction. 
The next dip in the conductance reflects the
presence of a resonance of type $(n,n)$ that interacts
strongly with only one of the neighbor resonances.
The following "S-type" Fano line 
is again superimposed on a broad peak
and indicates the presence of two strong interacting resonances with
the same symmetry in the lateral and transport directions.
For a quantitative analysis of the conductance we have to determine 
from the charge analysis within the dot region
the value interval of $V_d$ that corresponds
to the number of electrons found experimentally\cite{goeres}.

\section{Conclusions}

We have provided in this paper a systematic treatment of the 
conductance through a quantum dot strongly coupled to wide conducting 
leads via short quantum point contacts. The electronic transport through 
this type of dots is essentially a scattering process in a low confining
potential, which requires a direct solution of the
two-dimensional Schr\"odinger equation with a nonseparable
scattering potential. For this purpose, we have used a generalized 
scattering theory that allows for a complete description inside 
and outside the
scattering area and is based on the R-matrix formalism. The resonances 
are determined as poles of the
multidimensional scattering matrix, which contains the information about
channel mixing due to the nonseparable scattering potential.
The strong coupling of the quantum dot to the environment
yields overlapping resonances, which show a significant interaction 
with each other in the case of a favorable parity of the corresponding
resonant states.

The conductance is determined as a function of the potential energy within 
the dot region, and
every peak in conductance is associated with a resonance or a group of 
overlapping resonances. Based on the representation of the scattering matrix
in terms of the R-matrix 
we provide for each peak an exact decomposition of the conductance in 
resonant terms associated with each of the overlapping resonances and 
a background. The decomposition is hierarchical, i. e., from the 
strongest to the broadest resonance, and allows for a deep understanding 
of the phenomena, which determine the transmission in the case 
of interacting resonances.
The resonant states characterizing the open quantum dot in the Fano 
regime are presented in comparison with the eigenstates of 
the isolated dot. Every resonant state has a correspondent 
between these eigenstates  and, we distinguish between 
slight and strong modified states due to the coupling with the environment.
The last ones are called {\it hybrid resonant modes}, and 
they occur 
only in the case of a strong coupling regime of the quantum dot 
to the contacts, as an effect of the interaction between resonances
with the same parity. 
The phenomenon of hybridization evidenced here for
the quantum dots in the Fano regime of transport
attests the molecule-like behavior of this system and
opens the possibility to realize artificial 
molecules based on semiconductor nanostructures.

The conductance through the quantum dot in the Fano regime of transport is
also compared qualitatively 
to the experimental data reported in Ref. [\onlinecite{goeres}].

\begin{acknowledgments}
One of us (P.N.R.) acknowledge partial support
from German Research Foundation through SFB 787
and from the Romanian Ministry of Education and Research
through the Program PNCDI2, Contract number 515/2008.

\end{acknowledgments}

\appendix

\section{Poles of the $\tilde{\cal{S}}$-matrix}
\label{poles}

The starting point for our pole analysis is the expression of the
non-constant part in the $\tilde{\cal{S}}$-matrix in terms of
$\mathbf{A}_\lambda=\vec{\alpha}_\lambda \vec{\beta}^T_\lambda/(E-E_\lambda)$ 
and $\mathbf{\Omega}_\lambda$:
\begin{equation}
\mathbf{1} + i \mathbf{\Omega} = \left(\mathbf{1} + i \mathbf{A}_\lambda \right)
                           \left(\mathbf{1} + i \mathbf{\Omega}_\lambda \right).
\label{aux}
\end{equation}
Using the definition of $\vec{\beta}_\lambda$, Eq. (\ref{beta}),
we can immediately demonstrate that each determinant of the second order 
of $\mathbf{A}_\lambda$ is zero,
$ (\mathbf{A}_\lambda)_{ij} (\mathbf{A}_\lambda)_{lp}
  -(\mathbf{A}_\lambda)_{ip} (\mathbf{A}_\lambda)_{lj}=0$,
where each index $i,j,l,p$ is a composite index $(sn)$ with $s=1,2$, 
$n \ge 1$.
Therefore, the matrix $\mathbf{A}_\lambda$ has the rank 1. 
On this basis we find that
\begin{equation}
\mbox{det}\left[ \mathbf{1} + i \mathbf{A}_\lambda
          \right]
= 1 + i \mbox{Tr}[\mathbf{A}_\lambda].
\label{det}
\end{equation}
In order to demonstrate the above relation we consider a $M \times M$ 
matrix $\mathbf{A}$,
$M \ge 2$, with $\mbox{Rank}[\mathbf{A}]=1$,
calculate the determinant of $\mathbf{1}+ \mathbf{A}$ 
 and take after that
the limit $M \rightarrow \infty$.
According to the definition,  the determinant  
is given as a sum over all
permutations $\pi$ of the numbers $\{1,...,M\}$
\begin{eqnarray}
\mbox{det}\left[ \mathbf{1} + \mathbf{A} \right]
& = &
 \sum_{\pi_M} \mbox{sgn} \pi_M (\delta_{1m_1}+A_{1m_1})
                               ....
                               (\delta_{Mm_M}+A_{Mm_M}),
\end{eqnarray}  
with $m_1,...,m_M \in \{1,...,M\}$, $m_i \ne m_j$ for $i \ne j$,
$i,j \in \{1,...,M\}$ and $\mbox{sgn} \pi_M$ denotes the signature of the
permutation $\pi$ and $\delta_{ij}$ is the Kronecker delta. 
After a direct calculation we obtain
\begin{eqnarray}
\mbox{det}\left[ \mathbf{1} + \mathbf{A} \right]
& = &
 \;\;\; \sum_{\pi_M} \mbox{sgn} \pi_M \delta_{1m_1} ....\delta_{Mm_M}
\nonumber \cr
&   &
  +   \sum_{\pi_M} \mbox{sgn} \pi_M \delta_{1m_1} ....
                                    \delta_{M-1m_{M-1}}A_{Mm_M}
\nonumber \cr
&   &
  +   ........
\nonumber \cr
&   &
 +    \sum_{\pi_M} \mbox{sgn} \pi_M A_{1m_1} \delta_{2m_2} ....
                                    \delta_{Mm_M}
\nonumber \cr
&   &
 +    \sum_{\pi_M} \mbox{sgn} \pi_M \delta_{1m_1} .... \delta_{M-2m_{M-2}} 
                                    A_{M-1m_{M-1}}  A_{Mm_M}
\nonumber \cr
&   &
  +   ........
\nonumber \cr
&   &
 +    \sum_{\pi_M} \mbox{sgn} \pi_M A_{1m_1} ....  A_{Mm_M}
\nonumber \cr
& = & 1 + A_{MM} + ... + A_{11} 
     +\left| \begin{array}{cc} A_{M-1 M-1} & A_{M-1 M} \\
                               A_{M M-1}   & A_{M M}
             \end{array}
      \right| 
     + .... + \mbox{det}[\mathbf{A}].
\end{eqnarray}
Thus, $\mbox{det}\left[ \mathbf{1} + \mathbf{A} \right]$ is given as 1 plus 
a sum of determinants of different order of $\mathbf{A}$. But
$\mbox{Rank}[\mathbf{A}]=1$ and consequently all the determinants 
of $\mathbf{A}$ up to the second order are zero. So that we find
\begin{equation}
\mbox{det}\left[ \mathbf{1} + \mathbf{A} \right] = 1 + \mbox{Tr}[\mathbf{A}].
\label{aux4}
\end{equation} 
This result does not depend explicitly on the matrix dimension $M$ 
so that we can generalize
it for the case $M \rightarrow \infty$ and obtain Eq. (\ref{det}).

In the next step we calculate the adjugate matrix of
$\mathbf{1} + \mathbf{A}$, i. e., $\overline{\mathbf{1} + \mathbf{A}}$,  
in order to invert it: 
\begin{equation}
(\mathbf{1} + \mathbf{A})^{-1} =\frac{1}{\det[1+\mathbf{A}]}\overline{\mathbf{1} + \mathbf{A}}.
\end{equation}
For a given pair of indices $ij$ the corresponding matrix element of 
$\overline{\mathbf{1} + \mathbf{A}}$ is calculated as the product between $(-1)^{i+j}$
and the minor $ij$ of $(\mathbf{1} + \mathbf{A})^T$
(the determinant of the matrix obtained from
$(\mathbf{1} + \mathbf{A})^T$ by removing the row $i$ and the column $j$,
where $T$ denotes the matrix transpose).
Thus, for $i=j$ we obtain 
\begin{equation}
(\overline{\mathbf{1} +\mathbf{A}})_{ii} = \mbox{det}[\mathbf{1} +\mathbf{B}_{ii}],
\end{equation}
where $\mathbf{B}_{ii}$ is obtained from $\mathbf{A}^T$ by removing
the row $i$ and the column $i$. 
The matrix $\mathbf{B}_{ii}$ is a $(M-1) \times (M-1)$ 
matrix with the rank 1
and therefore 
\begin{equation}
(\overline{\mathbf{1} +\mathbf{A}})_{ii} = 1 + \mbox{Tr}[\mathbf{B}_{ii}]
                              = 1 + \sum_{j=1}^{M-1} (\mathbf{B}_{ii})_{jj}
                              = 1 + \mbox{Tr}[\mathbf{A}] - A_{ii}.
\label{adjoint1}
\end{equation}
In the case $j=i+1$ we find 
\begin{equation}
{\overline{(\mathbf{1} +\mathbf{A})}}_{i i+1} 
= -\mbox{det}[\mathbf{I}_i +\mathbf{B}_{ii+1}],
\label{aux1}
\end{equation}
where $(M-1) \times (M-1)$ matrix 
$\mathbf{I}_i$ is obtained from the unity matrix by changing 1 
on the position  $ii$ with 0, ($i < M$) and $\mathbf{B}_{i i+1}$ is the matrix
obtained from $\mathbf{A}$ by removing the $i$-th row and the $(i+1)$-th column. 
This is a $(M-1) \times (M-1)$ matrix
and has the rank 1. 
The matrix element $ii$ of $\mathbf{B}_{i i+1}$ is $A_{ii+1}$.
Further we write explicitly the determinant as a sum over all permutation
of the numbers $\{1,...,M-1\}$,
\begin{eqnarray}
\mbox{det}[\mathbf{I}_i +\mathbf{B}_{ii+1}]
& = &
 \sum_{\pi_{M-1}} \mbox{sgn} \pi_{M-1} (\delta_{1m_1}+b_{1m_1})
                                      ....
                                     (\delta_{i-1 m_{i-1}}+b_{i-1 m_{i-1}})
                                     b_{i m_i}
\nonumber \cr
&   & \qquad \qquad \qquad \times 
                                   (\delta_{i+1 m_{i+1}}+b_{i+1 m_{i+1}})
                                   ....
                                   (\delta_{M-1 m_{M-1}}+b_{M-1 m_{M-1}}),
\end{eqnarray}
where $b_{jl}$, with $j,l=\overline{1,M-1}$,
means the matrix element $jl$ of $\mathbf{B}_{ii+1}$.
Replacing $b_{i m_i}$ by $\delta_{i m_i}+b_{i m_i}-\delta_{i m_i}$
allows us to express 
$\overline{(\mathbf{1} + \mathbf{A})}_{i i+1}$
as $\mbox{det}[\mathbf{1} + \mathbf{B}_{ii+1}]$ minus the minor 
$ii$ of $\mathbf{I}_i + \mathbf{B}_{i i+1}$.
The last two determinants can be calculated using Eq. (\ref{det})
because the corresponding matrices are a sum of the unity matrix and a part
of $A$-matrix which has the property $\mbox{Rank}[\mathbf{A}]=1$.
Thus, we find
\begin{equation}
\mbox{det}[\mathbf{I}_i +\mathbf{B}_{ii+1}]
 = 1 + \sum_{j=1}^{M-1} b_{jj}
   -\left[1 + \sum_{j=1}^{i-1} b_{jj} + \sum_{j=i+1}^{M-1} b_{jj}
    \right]
\end{equation}
and after that, using Eq. (\ref{aux4}),
\begin{equation}
{\overline{(\mathbf{1} +\mathbf{A})}}_{i i+1} = -A_{i i+1}.
\label{adjoint2}
\end{equation}

Further we analyze the case $j > i+1$. If we eliminate the $i$-th row and
the $j$-th column in $(\mathbf{1}+\mathbf{A})^T$ we obtain a matrix which does
not have any more elements of the type $1+a_{ll}$ on the main diagonal
between the column $i$ and $j-1$. These elements are on the positions
$l-1l$, $l=\overline{i+1,j-2}$. Taking into account that we only 
need to calculate the determinant of this matrix we exchange the columns: 
$i \leftrightarrow i+1$, ..., $j-2 \leftrightarrow j-1$
and each of these $j-i-1$ operations changes the determinant with -1.
So that we can write
\begin{equation}
{\overline{(\mathbf{1} +\mathbf{A})}}_{i j}
= (-1)^{i+j+j-i-1} \mbox{det}[\mathbf{I}_{j-1} +\mathbf{B}_{ij}].
\label{aux2}
\end{equation}
As described above
the matrix $\mathbf{B}_{ij}$ is obtained from $\mathbf{A}^T$
by removing the $i$-th row and the $j$-th column and   
after that by exchanging the columns $i \leftrightarrow i+1$, ..., $j-2
\leftrightarrow j-1$. This $(M-1) \times (M-1)$ matrix has also 
rank 1 and  
the matrix element $j-1j-1$ of $\mathbf{B}_{ij}$ is $A_{ij}$, i. e.,
$\left(\mathbf{B}_{ij}\right)_{j-1 j-1}=A_{ij}$.
Similar to the previous case we can demonstrate here that
\begin{equation}
{\overline{(\mathbf{1} +\mathbf{A})}}_{i j} =-A_{ij}.
\label{adjoint3}
\end{equation}
With the same procedure we can demonstrate that the above result 
remains also valid for $j < i$.

If we put together the main results of this section, Eqs. (\ref{det}),
(\ref{adjoint1}), (\ref{adjoint2}), and (\ref{adjoint3}),
we find 
\begin{equation}
(\mathbf{1} + \mathbf{A})^{-1} = \mathbf{1} - \frac{\mathbf{A}}{1+\mbox{Tr}[\mathbf{A}]}.
\label{inv}
\end{equation}
The above relation does not depend essentially on $M$, so that we can 
take the limit $M \rightarrow \infty$ and generalize Eq. (\ref{inv})
for $i \mathbf{A}_\lambda$. After that we obtain from Eq. (\ref{aux}) 
that
\begin{equation}
(\mathbf{1} + i \mathbf{\Omega})^{-1} 
= (\mathbf{1} + i \mathbf{\Omega}_\lambda)^{-1}
  \left(\mathbf{1} - \frac{i \mathbf{A}_\lambda}
                       {1 + i \mbox{Tr}[\mathbf{A}_\lambda]}
  \right).
\label{inverse}
\end{equation}
Feeding this relation into the definition of
$\tilde{\cal{S}}$-matrix, Eq. (\ref{Stilde2}),  we find Eq.
(\ref{Stilde3}).

\bibliography{fano1}

\end{document}